\documentclass{emulateapj}

\newcommand{\vdag}{(v)^\dagger}
\newcommand\aastex{AAS\TeX}
\newcommand\latex{La\TeX}

\graphicspath{{./}{figures/}}

\shorttitle{CO$_2$ Reservoirs of Habitable Water-Worlds}
\shortauthors{N. Marounina and L. A. Rogers}

\begin{document}

\title{Internal Structure and CO$_2$ Reservoirs of Habitable Water-Worlds}

\email{nmarounina@gmail.com}

\author{Nadejda Marounina and Leslie A. Rogers}
\affil{Department of Astronomy and Astrophysics, University of Chicago,
    Chicago, IL 60637}



\begin{abstract}

 Water-worlds are water-rich ($>$~1~wt\%~H$_2$O) exoplanets. The classical models of water-worlds considered layered structures determined by the phase boundaries of pure water. However, water-worlds are likely to possess comet-like compositions, with between $\sim$~3~mol\% to 30~mol\% CO$_2$ relative to water. In this study, we build an interior structure model of habitable (i.e. surface-liquid-ocean-bearing) water-worlds using the latest results from experimental data on the CO$_2$-H$_2$O system, to explore the CO$_2$ budget and to localize the main CO$_2$ reservoirs inside of these planets. 
We show that CO$_2$ dissolved in the ocean and trapped inside of a clathrate layer can not accommodate a cometary amount of CO$_2$ if the planet accretes more than 11~wt\% of volatiles (CO$_2$~+~H$_2$O) during its formation.
We propose a new, potentially dominant, CO$_2$ reservoir for water-worlds: CO$_2$ buried inside of the high-pressure water ice mantle as CO$_2$ ices or (H$_2$CO$_3$~$\cdot$~H$_2$O), monohydrate of carbonic acid. 
If insufficient amounts of CO$_2$ are sequestered either in this reservoir or the planet's iron core, habitable zone water-worlds could generically be stalled in their cooling before liquid oceans have a chance to condense.

\end{abstract}



\section{Introduction}

Water-rich exoplanets ($>$1\% water by mass) will be the next accessible targets on the path of the observation of Earth-like planets \citep[e.g.][]{Beichman:2014}. Water-rich planets possess lower densities and larger radii than terrestrial Earth-like planets with similar masses, making them more amenable to observations and characterization by future surveys. \cite{Kuchner:2003} and \cite{Leger:2004fh} were the first to propose the existence of water-rich planets, suggesting that they would form beyond the snow line and accrete comet-like proportions of rocky material and ices. Then, interactions with protoplanetary disk or with other bodies in the system would bring them into the habitable zone (HZ) of their star, forming a global water ocean at their surfaces. Since then, \citet{Luger:2015de} showed that photo-evaporation of a H$_2$/He envelope from a mini-Neptune could be another path of formation of water-worlds, especially relevant for planets in the HZs of M-dwarfs. Additionally, several theoretical studies predict an efficient accretion of volatiles during planet assembly (especially in scenarios with low-mass host stars and long-lived protoplanetary disks), forming planets with up to $\sim$50\% of water by mass \citep{Raymond:2004de, Alibert:2017, Kite:2018}. Thus, water-rich planets are possibly numerous around M dwarfs and will be prime targets for atmospheric characterization in the near future. Indeed, the recent discovery and characterization of the TRAPPIST-1 system has shown that it may contain HZ planets with several tens of percent of water by mass \citep{Gillon:2017fw, Unterborn:2018ep, Grimm:2018, Unterborn:2018gr}.

At the present time, a widely accepted terminology for the denomination of water-rich exoplanets does not exist, and terms such as ``water-world" or ``ocean planet" may have different meanings from paper to paper. Here, we choose to call ``water-world" any planet with sufficient quantities of volatiles that it could form a high-pressure water ice layer given a favorable interior temperature profile (regardless of whether the planet actually cools sufficiently for the high-pressure ice to form). By this definition, a water-world could have a subsurface ocean, a surface global or partial ocean, or in the most extremes cases, no ocean at all with the volatile-rich envelope entirely in a supercritical or vapor state.

The common definition of the HZ does not apply to water-worlds, and the habitability of water-worlds is still poorly constrained. Computations of the HZs mostly focus on the Earth-like planets, where continent and seafloor weathering stabilizes the concentration of CO$_2$ in the atmosphere, inducing a negative feedback  \citep[e.g.][]{Walker:1981uq, Kasting:1993, Kasting:2003bp, Kopparapu:2013}. Such feedback is probably not active for habitable water-worlds, or water-worlds with global surface water oceans, because silicates are likely to be isolated from liquid water by a high-pressure ice mantle \citep{Leger:2004fh, Sotin:2007fh, Fu:2010ch, Levi:2013eq, Levi:2017gv}. Assuming that the high-pressure ice layer precludes chemical exchanges between the silicate layers and liquid water, the partial pressure of CO$_2$ in the atmosphere would be controlled by the total amount of CO$_2$ in the hydrosphere (i.e. volatile-dominated layers of the planet) and the temperature profile inside of these volatile-rich layers.

Water-worlds are likely to possess CO$_2$-rich bulk compositions.  \citet[][]{Kuchner:2003, Leger:2004fh, Selsis:2007jx} proposed that water-worlds would initially form with comet-like compositions. In comets, CO$_2$ is the second most abundant volatile after H$_2$O, ranging from $\sim$~3~mol\% to 30~mol\% relative to water \citep[e.g.][]{Bockelee:2004, Mumma:2011jn, Ootsubo:2012cg}. For these planets, the pressure at the ice-silicate interface exceeds the limiting pressure for volcanic degassing \citep[$\sim$~0.6GPa,][]{Kite:2009kv}. Therefore, CO$_2$ is not supplied to the hydrosphere by volcanism from the silicate layers of the planet, and the total CO$_2$ mass in the hydrosphere does not vary substantially after the planet forms and differentiates.

In the scenario where water-worlds form by photo-evaporation \citep{Luger:2015de}, the amount of CO$_2$ left in the planet's atmosphere would depend on the overall CO$_2$ accretion and escape history. \citet{Luger:2015de} propose that planets that have lost their hydrogen envelopes may still possess high-density atmospheres with considerable amounts of CO$_2$.

Only a small fraction of the CO$_2$ accreted by a water-world can reside in the planet's atmosphere if the water-world is to maintain a temperate surface temperature and a surface liquid water ocean. Indeed, even the highest CO$_2$ partial pressures allowed inside of the HZ (P$_{CO_2}\sim$100~bar, Fig. \ref{fig:HZ}) correspond to less than ~1~wt\% CO$_2$ relative to the total volatile content of the planet.
 Depending on the orbital separation of the planet, the host star, and the planet history, higher amounts of CO$_2$ in the atmosphere could lead either (i) to the evaporation of the liquid oceans or (ii) to the prevention of liquid oceans from condensing in the first place as the planet is impeded from cooling from its post-accretion hot state. To avoid this situation, much of the CO$_2$ accreted by the planet would need to be sequestered in the planetary interior, separate from the atmosphere because CO$_2$ reservoirs in contact with the atmosphere could be easily destabilized by temperature changes \citep{Kitzmann:2015, Levi:2017gv}.

The water-dominated layers of the hydrosphere can store only a limited amount of CO$_2$. If these layers are \textit{saturated}, the \textit{excess} of carbon dioxide would form a new, separate phase. To set an upper limit on the amount of CO$_2$ that can be stored in these water-rich layers, here we consider the extreme scenario in which they are fully saturated. Therefore, in the rest of this work, we distinguish between CO$_2$ \textit{saturated} reservoirs (which designate the storage of CO$_2$ by saturating water-dominated layers), and \textit{excess} CO$_2$ reservoirs (i.e. CO$_2$ reservoirs that would form if water-dominated layers are saturated). One example of a saturated reservoir would be the water ocean, while an example of an excess reservoir is the atmosphere.

Previous work on water-rich exoplanets explored the potential interactions between the ocean and the atmosphere in great detail, but has not examined how/if the overall CO$_2$ budget and the distribution of CO$_2$ throughout the hydrosphere would even allow the presence of liquid water at the surface of water-worlds. \citet{Kitzmann:2015} was the first to compute the HZ of water-worlds possessing liquid water oceans at their surfaces. The study constrains the size of the HZ for a range of partial pressures of CO$_2$. It also points out that a solubility-controlled CO$_2$ abundance in the atmosphere constitutes an unstable CO$_2$ feedback cycle and any perturbation in temperature or atmospheric CO$_2$ content could lead to a runaway greenhouse or the freezing of the surface. The extensive study of \citet{Levi:2017gv}  proposed a mechanism to stabilize CO$_2$ partial pressure in the atmospheres of water-worlds by accounting for a wind-driven surface circulation and sea-ice formation at higher latitudes \citep[see also][]{Ramirez:2018wn}. To date, hydrosphere structures, CO$_2$ contents and CO$_2$ reservoirs in the interiors of these CO$_2$-rich water-worlds are poorly constrained, and this is what we aim to study here.
 
Here, we build a planet interior structure model using the latest results from experimental data on the CO$_2$-H$_2$O system to explore the CO$_2$ budget and to localize the main CO$_2$ reservoirs inside of water-worlds. The classical models of water-worlds \citep{Kuchner:2003, Leger:2004fh, Selsis:2007jx} considered layered structures determined by the phase boundaries of pure water. As a first application of our new water-world interior structure model, we perform the thought experiment of considering complete saturation in CO$_2$ of these water-dominated layers. By quantifying the maximum amount of CO$_2$ possibly stored inside each of these water-rich layers, we assess whether cometary amounts of CO$_2$ can be accommodated in the classical structure models for water-worlds. 

Section~\ref{sec:Model} describes the thermodynamic and planetary model that we developed. We apply the model to quantify the potential saturated and excess CO$_2$ reservoirs in the hydrospheres of water-worlds in Sections~\ref{sec:ResultsSaturated} and ~\ref{sec:ResultsExcess}, respectively. We discuss the limitations of our model and uncertainties of the current equations of state in Section~\ref{sec:Discussion} and summarize our conclusions in Section~\ref{sec:Conclusion}.


\section{Model}
\label{sec:Model}

\begin{figure}
\centering
\includegraphics[width=1\linewidth]{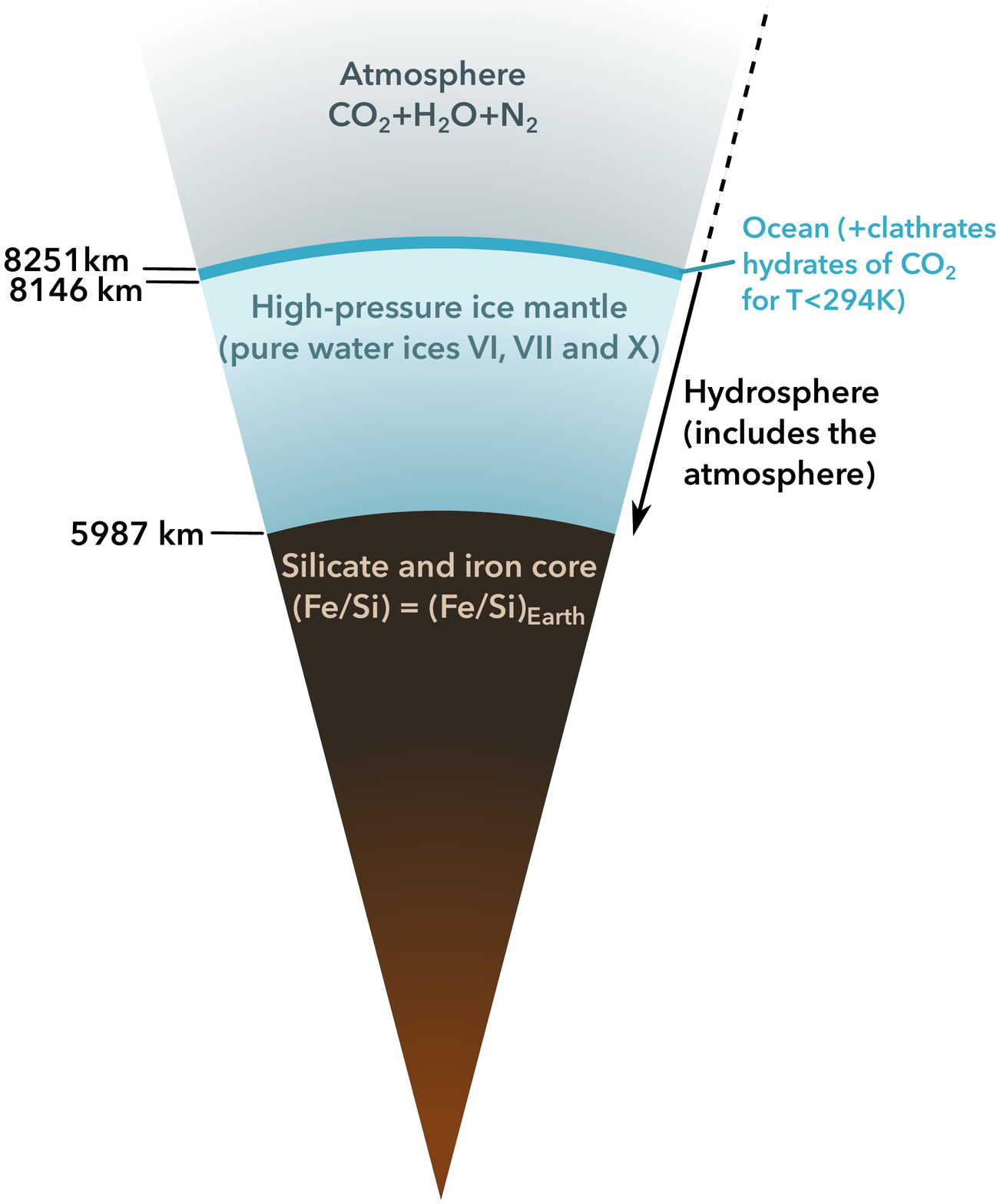}

\caption{\label{fig:struct} An example of an interior structure of a water-world, for 33~\% of water by mass and an isotherm of 300~K in the ocean. Layer thicknesses are true to scale.}
\end{figure}

We consider H$_2$O-and-CO$_2$-rich, fully differentiated planets. The internal structure of these planets is differentiated into a rocky core with a roughly Earth-like Fe/silicate ratio surrounded by upper volatile-rich layers, which we call the ``hydrosphere" (c.f. Fig. \ref{fig:struct}). Our study focuses on the structure of the hydrosphere. 

\begin{figure}
\centering
\includegraphics[width=1\linewidth]{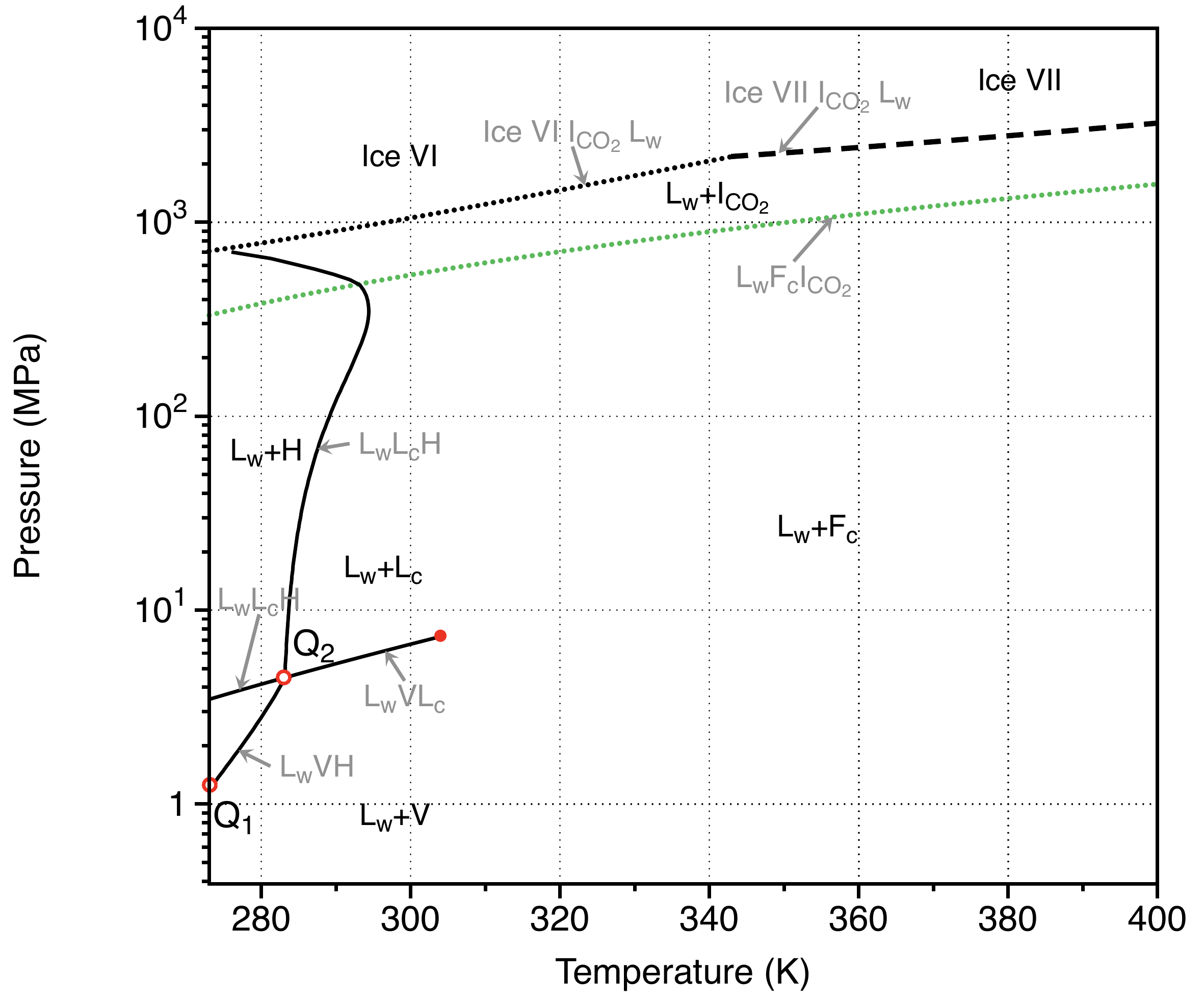}

\caption{\label{fig:phasediagram} P-T projection of the CO$_2$-H$_2$O phase diagram, showing the two- and three-phase coexistence curves and CO$_2$ critical point. Annotations specify the various phases including the vapor phase V, water-rich liquid L$_w$, CO$_2$-rich liquid L$_c$, fluid CO$_2$ phase (above the critical temperature of CO$_2$) F$_c$, hydrate phase H, CO$_2$ ice I$_{CO_2}$ and ices VI and VII. The filled red circle denotes the critical point of CO$_2$, while the open red circles Q$_1$ and Q$_2$ are quadruple points where there is a coexistence of liquid water, vapor, hydrate and water ice Ih (Q$_1$) and liquid water, vapor, hydrate, and liquid CO$_2$ (Q$_2$). The ice curves are from \citet[][]{Abramson:2017ie}, the clathrate curve is from TREND (see the comparison with the experimental data Fig. \ref{fig:comp_trend_exp}), and the values of the quadruple points and the triple lines that cross them are from \citet{Wendland:1999jy}. }
\end{figure}

Depending on the temperature and pressure profile in the hydrosphere, phases such as high-pressure ice, clathrate hydrates of CO$_2$, solid or liquid CO$_2$, liquid water or gas could form (Fig. \ref{fig:phasediagram}). We compute the partitioning of water and CO$_2$ between these phases for a range of planetary masses ($M_p$), volatile (H$_2$O+CO$_2$) mass fractions respective to the total planet mass ($X_v=M_v/M_p$, where $M_v=M_{H_2O}+M_{CO_2}$ is the total volatile mass), mass fraction of CO$_2$ relative to the total volatile mass ($X_{CO_2}=M_{CO_2}/M_v$), and an assumed temperature-pressure profile. As a result, we obtain an internal structure of the hydrosphere, or the separation of the hydrosphere in several layers assuming thermodynamic equilibrium. For this study, the silicate and iron parts of the planet contribute to the model only in setting the mass-radius boundary conditions at the base of the hydrosphere.

As we explore planets with surface water oceans, the surface temperature (i.e., the temperature at the ocean-atmosphere interface) must be between 273.15~K and 400~K. For surface temperatures lower than 273.15~K, the ocean has an icy crust at its surface, which would make the liquid water ocean challenging to detect by current and planned surveys. At the other extreme, $\sim$400~K is the highest temperature for the life as we know it \citep{Holden:2010ep, Corkrey:2014cp}.

We focus on planets with sufficient water to form high pressure ice mantles, wherein reactions between the silicate rocks and liquid water are suppressed and assumed to be negligible. This assumption is commonly made in the study of water-worlds \citep[e.g.][]{Fu:2010ch, Kitzmann:2015, Levi:2017gv}. The absence of significant liquid water-rock chemical interactions has two important consequences for the internal structure and CO$_2$ reservoirs of water-worlds. First, though formation of carbonates --- the entrapment of CO$_2$ as (Ca, Mg, Fe)CO$_3$ --- is an important carbon reservoir on the Earth, it is negligible for water-worlds. Second, in the absence of additional chemical agents that could drive the pH (i.e. salts) and increase the speciation of CO$_2$, we can use the equation of state for pure H$_2$O-CO$_2$ mixtures to model water-worlds. We further discuss and quantify the effect of water-rock interactions in \S~\ref{sec:CO2vsRocks}. 

Fig.~\ref{fig:phasediagram} shows the possible phases of CO$_2$+H$_2$O mixtures over the temperature and pressure ranges relevant to habitable water-worlds. Areas between lines represent ranges of temperatures and pressures where one phase can exist, or where two phases are in equilibrium. Lines trace three-phase or two-phase equilibria. Lastly, points mark both specific combinations of temperature and pressure where four phases are in equilibrium (also called quadruple points) and the critical point of CO$_2$. Vapor (V) and water-rich liquid phases (L$_w$) coexist in equilibrium at the low pressures end of this diagram. At higher pressures (P$\gtrsim$1.2~MPa) and low temperatures (T$<$294~K), CO$_2$+H$_2$O mixtures form a phase called clathrate hydrate. Clathrate hydrate is a crystalline structure consisting of a lattice of water molecules, organized as cages entrapping guest molecules (in this case CO$_2$). For the structure to be stable, guest molecules have to occupy a minimum fraction of the water cages. The total occupancy of the cages varies with temperature and pressure \citep{Sloan:2007cl}. Clathrate hydrates naturally occur on Earth \citep{Sloan:2007cl} and their presence has been hypothesized on Mars \citep[e.g.][]{Kite:2017bz} and icy satellites \citep{Tobie:2006fj, Choukroun:2010ca}.

For even higher pressures (P$\gtrsim$4.5~MPa), CO$_2$ may condense in a liquid (noted as L$_c$ in Fig. \ref{fig:phasediagram}). Under these conditions, instead of a vapor-liquid equilibrium, we have a liquid-liquid equilibrium with a water-rich liquid and a CO$_2$-rich liquid. For the highest range of pressures explored in this study (480~MPa and above), CO$_2$ ice (I) and polymorphs of high-pressure water ice form.

\subsection{Atmosphere}
We use the 1-D radiative-transfer/climate model CLIMA, originally developed by \citet{Kasting:1986ix}, and most recently updated in calculations of the HZs of Earth-like exoplanets \citep{Kopparapu:2013, Kopparapu:2014}. CLIMA uses a correlated-k method to calculate the absorption coefficients of spectrally active gases both for the incoming shortwave stellar radiation (in 38 solar spectral intervals ranging from 0.2 to 4.5~$\mu$m) and for the outgoing longwave IR radiation (in 55 spectral intervals spanning wavenumbers from 0 to 15,000~cm$^{-1}$). The 2-stream multiple scattering method of \citet{Toon:1989} is used to calculate the radiative heating rate in each of the 101 atmospheric layers. 
For the inner edge of the HZ, the assumed atmospheric pressure-temperature profile consists of a moist pseudoadiabat extending from the surface up to an isothermal (200 K) stratosphere as described in \citet[Appendix A,][]{Kasting:1988gd}. For the outer edge of the HZ, a moist H$_2$O adiabat is assumed in the lower troposphere, and when condensation was encountered in the upper troposphere a moist CO$_2$ adiabat is used, as described in Appendix B of \citet{Kasting:1991}.

In this study, we focus on ``habitable" water-worlds or water-words with liquid water at their surfaces. In these cases, gases in the atmosphere are dissolved in the ocean. To place an upper limit on volatile content of \textit{saturated} reservoirs, we assume the atmosphere species are in thermodynamic phase equilibrium with the liquid water layer. Consequently, at the interface between the ocean and the atmosphere, the temperature, pressure and chemical potentials of all of the chemical species constituting these layers are equal. 

While CLIMA handles the calculation of the atmospheric radiative transfer and pressure - temperature profile in our model of water-worlds, phase changes and phase equilibria occurring in the fluid and solid water-rich layers of the hydrosphere are computed using TREND~3.0 software \citep{TREND:2016wr}.

\subsection{Equation of State for CO$_2$-H$_2$O Mixtures}

To estimate the potential CO$_2$ reservoirs of HZ water-rich exoplanets, we need to model the behaviour of CO$_2$-H$_2$O mixtures at temperatures corresponding to ``habitable" planet surfaces  (273~K- 400~K) and at pressures up to the  formation of high-pressure ice ($\sim$ 3.2~GPa). For this temperature range and at low to moderate pressures (up to few MPa), there are numerous equations of state for the H$_2$O-CO$_2$ system that compute both the behavior of individual phases and phase equilibria  \citep[e.g.][]{Carroll:1991, Spycher:2003fb, Hu:2007ca}. However, computing the behavior of CO$_2$+H$_2$O mixtures up to several GPa is currently still a challenge due to the lack of experimental data in the region of pressures above  $\sim$1~GPa. The recent study of \citet{Abramson:2017ie} provides the first experimental data in this high-pressure region.

\begin{figure}
     \centering
     \includegraphics[width=1\linewidth]{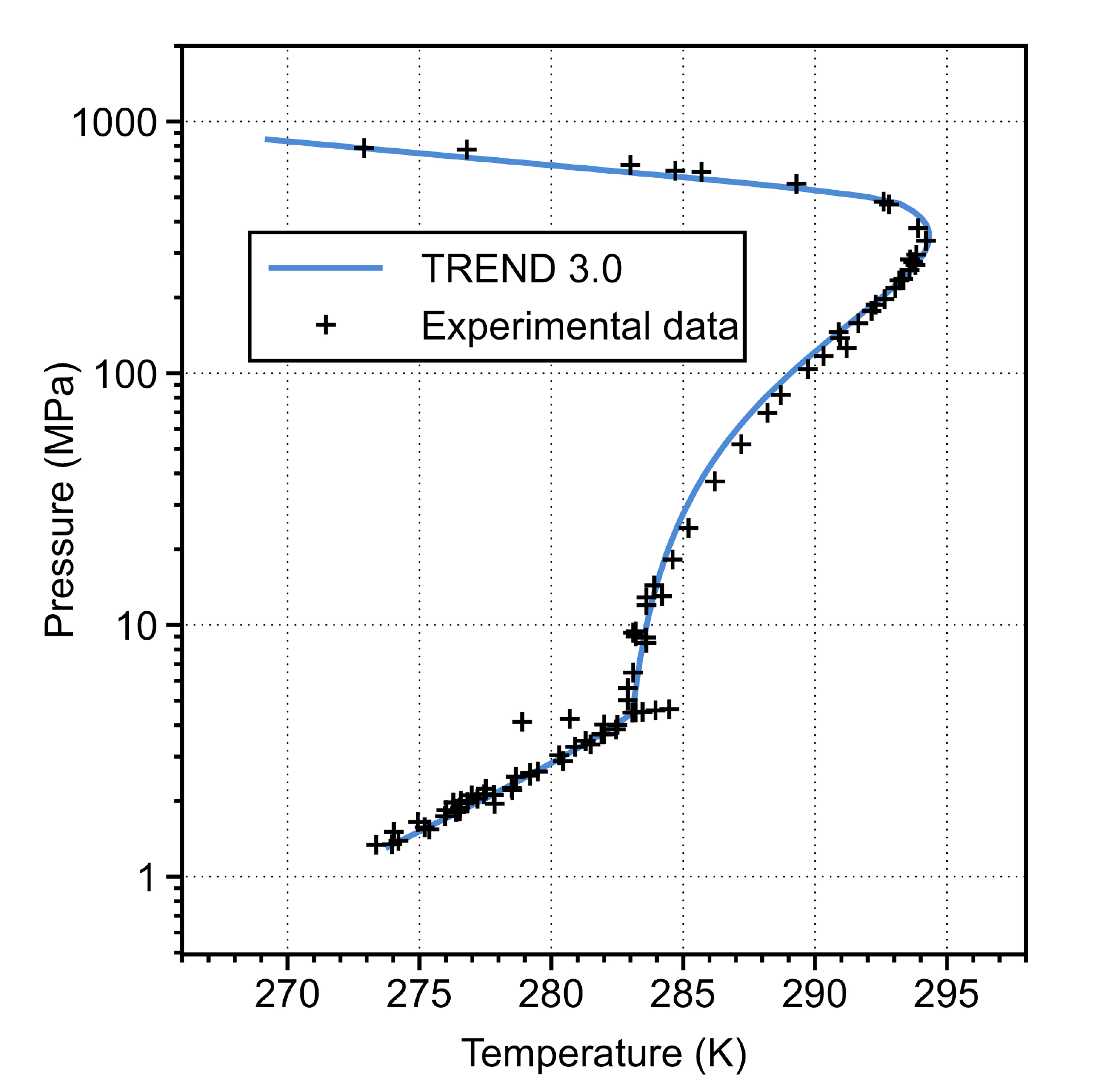}
     \caption{Comparison between the prediction of TREND 3.0 and the experimental data for the triple line CO$_2$ clathrates/liquid water-rich phase/CO$_2$-rich fluid phase. Experimental data are summarized in \citet{Sloan:2007cl}.}
     \label{fig:comp_trend_exp}
 \end{figure}

 We use the state-of-the-art equations of state (EOS) TREND~3.0 to model H$_2$O-CO$_2$ mixtures. TREND~3.0 is the reference EOS in the carbon capture, storage and transport industry, and is (to our knowledge) the best option to reproduce the behaviour of CO$_2$-H$_2$O system in the temperature and pressure ranges of relevance to ocean-bearing water-worlds \citep{Gernert:2016bc}. TREND~3.0 reproduces all of the available experimental data of CO$_2$-H$_2$O system for pressures up to 100~MPa with low errors (typically lower than 2~\%, see e.g. Fig. \ref{fig:comp_trend_exp} or detailed comparisons in \citet{Gernert:2016bc}). TREND~3.0 is the result of decades of development by specialists in fluid behaviour and construction of equations of state. We describe TREND~3.0 in detail below.

The model for CO$_2$-H$_2$O mixtures implemented in TREND~3.0 uses two empirical equations of state for pure fluids, explicit in Helmholtz free energy: one for water \citep[IAPWS,][]{Wagner:2002in} and one for CO$_2$ \citep{Span:1996cg}. Provided density $\rho$ and temperature $T$ as independent variables, these equations of state compute ideal and the residual parts of the Helmholtz energy and their first, second and third derivatives. By combining these derivatives, all thermodynamic properties can be calculated, including thermodynamic potentials $u$ (specific internal energy), $h$ (specific enthalpy), $g$ (specific Gibbs free energy) or $p$ (pressure), $s$ (specific entropy), and $C_p$ (heat capacity at constant pressure), to cite few of them. See the complete list of thermodynamic properties in Tab. 6.3 of \citet[][]{Wagner:2002in} and Tab. 3 of \citet[][]{Span:1996cg}.

The range of temperature-pressure-compositions where reliable experimental data exists defines the initial range of validity of the pure water and pure CO$_2$ equations of state. For pure water, IAPWS is valid up to 1273~K and 1~GPa, and for pure CO$_2$ the primary range of validity of the equation of state goes up to 1100~K and 800~MPa. However, special care has been taken to ensure that these equations of state yield reasonable results when extrapolated beyond this initial range of validity. The density of water predicted by extrapolating IAPWS up to 3.5~GPa deviates from the experimental data of \citet{Wiryana:1998kd} by no more than $3.5\%$. For CO$_2$, the equation of state of \citet{Span:1997} reasonably describes the behavior of the pure substance along the Hugoniot curve up to the limits of the chemical stability of carbon dioxide \citep[see Fig. 36 of][]{Span:1996cg}. Moreover, both of these equations of state perform well when compared to ideal curves up to very high pressures and temperatures. Ideal curves  \citep[as defined by][]{Span:1997} are curves along which one property of a real fluid is equal to the corresponding property of the hypothetical ideal gas at the same temperature and density. Comparison to ideal curves has been shown to reliably predict the quality of extrapolations of empirical equations of state for pure substances \citep[][]{Span:1997, Deiters:1997fi, Span:2013}. 

To combine the equations of state of pure compounds and compute the properties of various phases of CO$_2$-H$_2$O mixtures, TREND~3.0 blends multiple approaches that we detail in the following paragraphs. The overall methodology of TREND~3.0 is to compute the Gibbs energy of the CO$_2$-H$_2$O mixture at a given $p$ and $T$ and overall molar composition $z$, and then to minimize it. TREND~3.0 then determines: which phases are stable at the provided $T$, $p$ and $z$; the density of all stable phases; and the molar composition of each of these phases. Once the above parameters are evaluated, TREND~3.0 computes the thermodynamic and calorific properties of each phase. For details of the algorithms used in TREND~3.0 refer to \citet{Kunz:2007fq}.

To compute the behavior of the CO$_2$-H$_2$O mixture from the pure fluid equations of state, TREND~3.0 uses mixing rules from \citet{Kunz:2007fq} as adapted in \citet{Gernert:2016bc}. The mixing parameters are derived by fitting the experimental data of the CO$_2$-H$_2$O system, that includes the small amount of CO$_2$ dissociated in HCO$_3^-$ and CO$_3^{2-}$. Consequently, the resulting equation of state that is  fitted to this data implicitly accounts for this dissociation, in absence of other solutes.

TREND 3.0 is widely used in the carbon capture and storage (CCS) community and is regularly updated upon the arrival of new experimental data \citep[e.g.][]{Kunz:2007fq, Kunz:2012fd, Lovseth:2018gn}. The model described in \citet{Gernert:2016bc}, EOS-CG, is based on the mathematical structure introduced in the GERG-2004 and GERG-2008 approaches \citep{Kunz:2007fq, Kunz:2012fd} but is explicitly developed to provide an accurate description of the CO$_2$-H$_2$O mixture. The initial range of validity of EOS-CG extends up to 500~K and 0.1~GPa;  within these ranges, the carefully vetted experimental data sets on CO$_2$-H$_2$O mixtures to which EOS-CG is fit agree to within $\pm$~2\% on the density of the mixture and to within $\pm$~0.3~mol\% on the solubility of CO$_2$ in water. \citet{Gernert:2016bc} could not extend the EOS-CG model fit to higher pressures and temperatures due to significant disagreements in the density and/or solubility measurements between experimental data sets; no data set could be identified as significantly more accurate than the others. Tests have shown that this mixing rule can reasonably be used outside the extended range of validity if larger uncertainties are acceptable \citep{Kunz:2012fd}. An estimate of the error introduced by extrapolating EOS-CG above $>$~0.1~GPa is shown in Fig. \ref{fig:profil}~D. 

 EOS-CG successfully describes the fluid phase that is in equilibrium with dry ice or CO$_2$-hydrate \citep{Gernert:2016bc}. TREND~3.0  detects the formation of CO$_2$ ice I using the model detailed in \citet{Jager:2012jo}, which proposes a thermal equation of state for solid carbon dioxide that is explicit in Gibbs energy. The initial range of validity of this equation of state is 80~K~$<$~T~$<$~300~K and 0~MPa~$<$~T~$<$~500~MPa. However, even when compared to experimental data with pressures up to 10~GPa \citep[for a 296~K isotherm][]{Liu:1984jg, Olinger:1982kx}, this equation of state reproduces the density of solid CO$_2$ within 3~\%.

 The formation of CO$_2$ clathrate hydrate is computed using the model described in a series of three papers: \citet{Vins:2016kx, Vins:2017hv} and \citet{Jager:2016dr}. The model is inspired by \citet{Sloan:2007cl} and based on the statistical van der Waals and Plattew approach \citep{Vdwaals:1959cs}. It computes the chemical potential of water in the hydrate lattice, $\mu^H_w$. Typically, if this chemical potential is lower than the chemical potential of liquid water, computed with EOS-CG, then the clathrate hydrate phase is stable.
 
 \citet{Vins:2016kx} has shown that the lattice parameter of hydrates and the Langmuir constants that quantify the molecular interactions between the water lattice and CO$_2$ guest molecule, strongly influence the computed value of $\mu^H_w$. The water lattice parameter of clathrate hydrates is especially important at high pressures, while the Langmuir constant affects the computed filling fraction of the clathrate cages, and therefore for the amount of CO$_2$ trapped in the clathrate layer. \citet{Vins:2017hv} describe the fitting procedure to obtain all the necessary model parameters. \citet{Jager:2016dr} provides the result of the fitting, assesses the performance of this model and details its implementation in the TREND~3.0. For the CO$_2$-H$_2$O mixture, this new model performs better when compared to previous formulations \citep[e.g.][]{Ballard:2002a, Ballard:2002b}. This model is valid for the whole range of temperature, pressure, and compositions of interest in this study (T=$\left[273K,400K\right]$, P up to $\sim$700~MPa, see Fig.~\ref{fig:comp_trend_exp}).
 
 TREND~3.0 does not account for the formation of high-pressure water ice. To detect the phase boundary of high-pressure water ice in contact with CO$_2$ ice, we use the scaling laws of \citet{Abramson:2017hl}, for ice VI and VII. In that study, the authors obtained and fitted experimental data for the triple line CO$_2$ ice/high-pressure water ice (either VI or VII, depending on the temperature)/liquid water. This triple line shows considerable deviations from the pure water system (e.g., a difference of 0.22~GPa for T~=~363~K). Thus water planet interior structure models that rely on the phase boundaries of pure water could incur significant errors.
 
 For the equation of state of ices VI, VII, and X (for the density and adiabatic temperature profiles of these phases) we use the formulation and the parameters summarized in \citet{Noack:2016bh}.

 \subsection{Water-World Interior Structure}

We use a planet interior structure model to  self-consistently compute the radii and hydrosphere structure of water-worlds (Fig~\ref{fig:struct}).

As inputs to the model, we specify the mass of the planet's rocky core M$_{rock}$ (assumed to have a Earth-like silicate to iron mass ratio), the planet's volatile mass $M_{v}=M_p-M_{rock}=X_v M_p$, and surface temperature. We assume an isothermal temperature profile in the liquid and clathrate layers, and an adiabatic temperature profile in the high-pressure ice layers  \citep{Fu:2010ch, Noack:2016bh}.

To model a planet, we guess an initial planet radius (defined at the ocean-atmosphere boundary). Then, we integrate the equations of hydrostatic equilibrium and the mass in a spherical shell inwards through the hydrosphere toward the center of the planet (with care to use the equation of state of the appropriate phase). We continue integrating inward until the total mass balance of volatiles is satisfied (i.e., when the integrated mass of the hydrosphere equals M$_{v}$). We then compare the radius at the water-silicate boundary obtained from the integration, R$_{wsb}$, to the radius R$_{rock}$ of an Earth-composition core of mass M$_{rock}$ under the pressure overburden of the volatile envelope. R$_{rock}$ is derived by interpolating the models for rocky cores from \citet{Rogers:2011gz}.
If R$_{wsb}$ and R$_{rock}$ are too disparate, we adjust the radius of the planet at the ocean/atmosphere interface and repeat the inward integration through the hydrosphere. We iterate until R$_{wsb}$ and R$_{rock}$ correspond with an error less than 0.1\%.
 
By modeling isothermal oceans saturated with CO$_2$ we set a strict upper limit on the CO$_2$ mass dissolved in water and trapped in clathrates. Adiabatic temperature profiles are associated with convection. Convection would lead to mixing and evolution towards a near-constant concentration of CO$_2$ through the ocean, mediated by the CO$_2$ partial pressure in the atmosphere. Because the solubility of CO$_2$ increases with pressure (Section~\ref{sec:profil}), the solubility of CO$_2$ in the liquid ocean is at its lowest at the ocean-atmosphere interface. Consequently, convecting adiabatic oceans would lead to lower CO$_2$ contents for the planet than those estimated with isothermal saturated ocean profiles. Further, an adiabatic temperature profile within the clathrate layer would lead to thinner clathrate layers, because of the increase in temperature with depth. This would in turn diminish the clathrate layer's capacity as a CO$_2$ reservoir.

There is also physical motivation to consider non-convecting oceans on water-worlds. \citet{Levi:2017gv} showed that oceans in water-worlds might not possess a deep overturning circulation, due to the lack of an energy source. The circulation in Earth's ocean relies on winds, lunar and solar tidal forcing, and the subsequent turbulent dissipation of internal waves on the seafloor topography. Water-worlds do not have strong topography at the oceanic floor, and the depth of the ocean of these planets is considerably more important than the Earth's ocean, requiring significant energy sources for a steady state global circulation. \citet{Levi:2017gv} show that winds on water-worlds could provide enough energy for convection inside of a $\sim$1-km surface layer. If the internal heat flux on these planets is low \citep{Levi:2014bc}, then the temperature profile inside the ocean will be conductive.


\section{CO$_2$ content of saturated reservoirs}
\label{sec:ResultsSaturated}

\subsection{Description of saturated reservoirs of water-worlds}
\label{sec:profil}
 
\begin{figure*}[h!]
    \centering
    \includegraphics[width=0.24\linewidth]{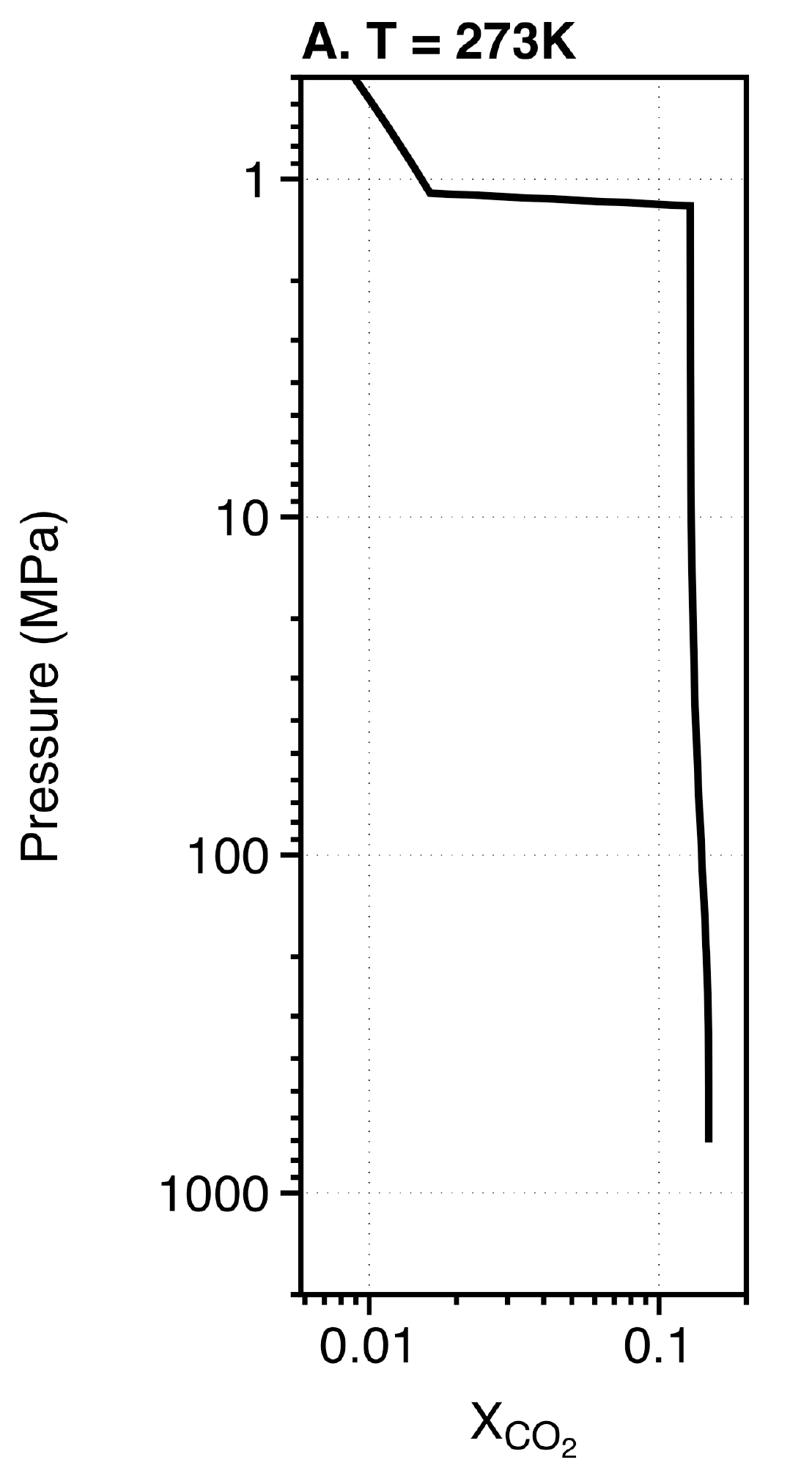}
    \includegraphics[width=0.24\linewidth]{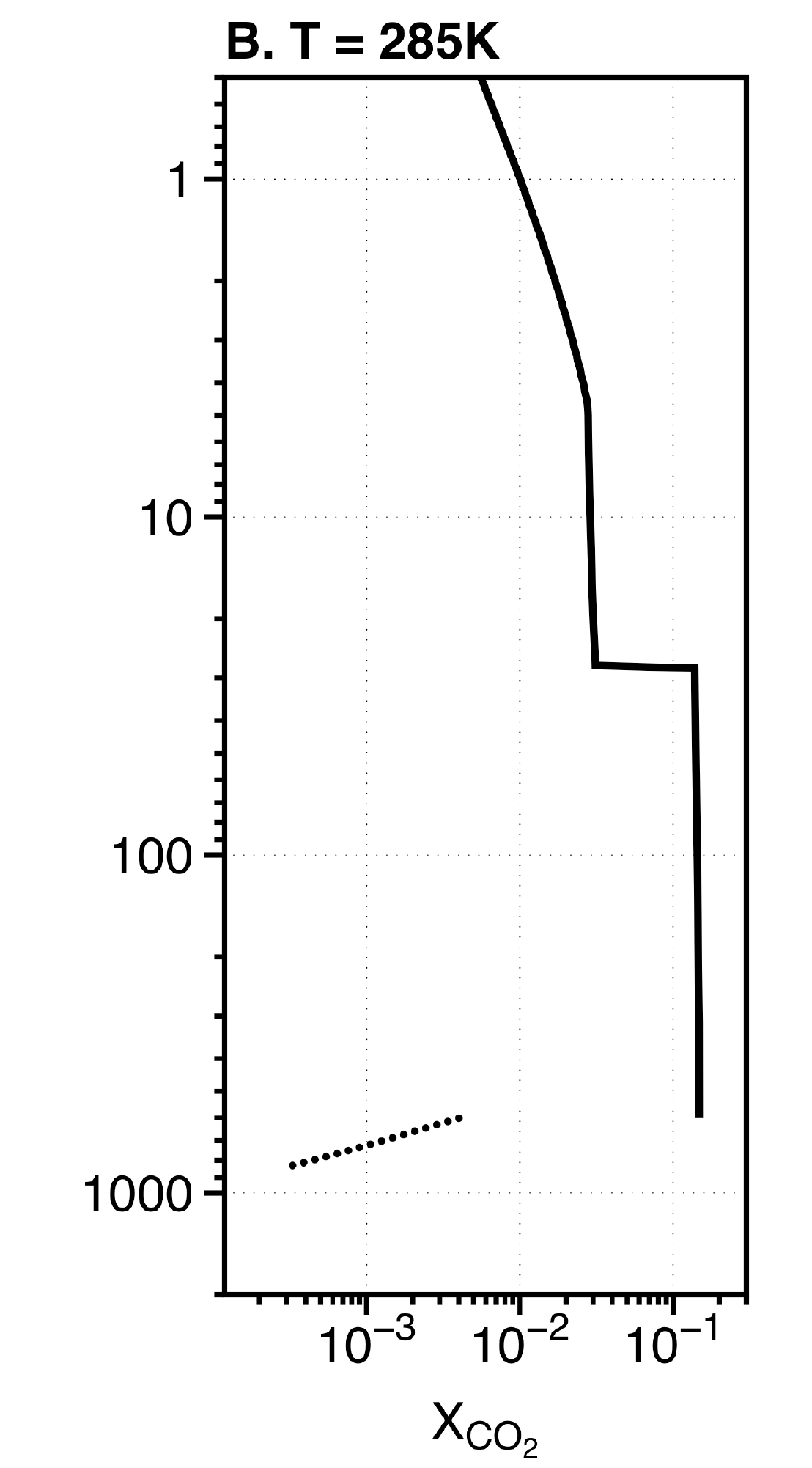}
    \includegraphics[width=0.24\linewidth]{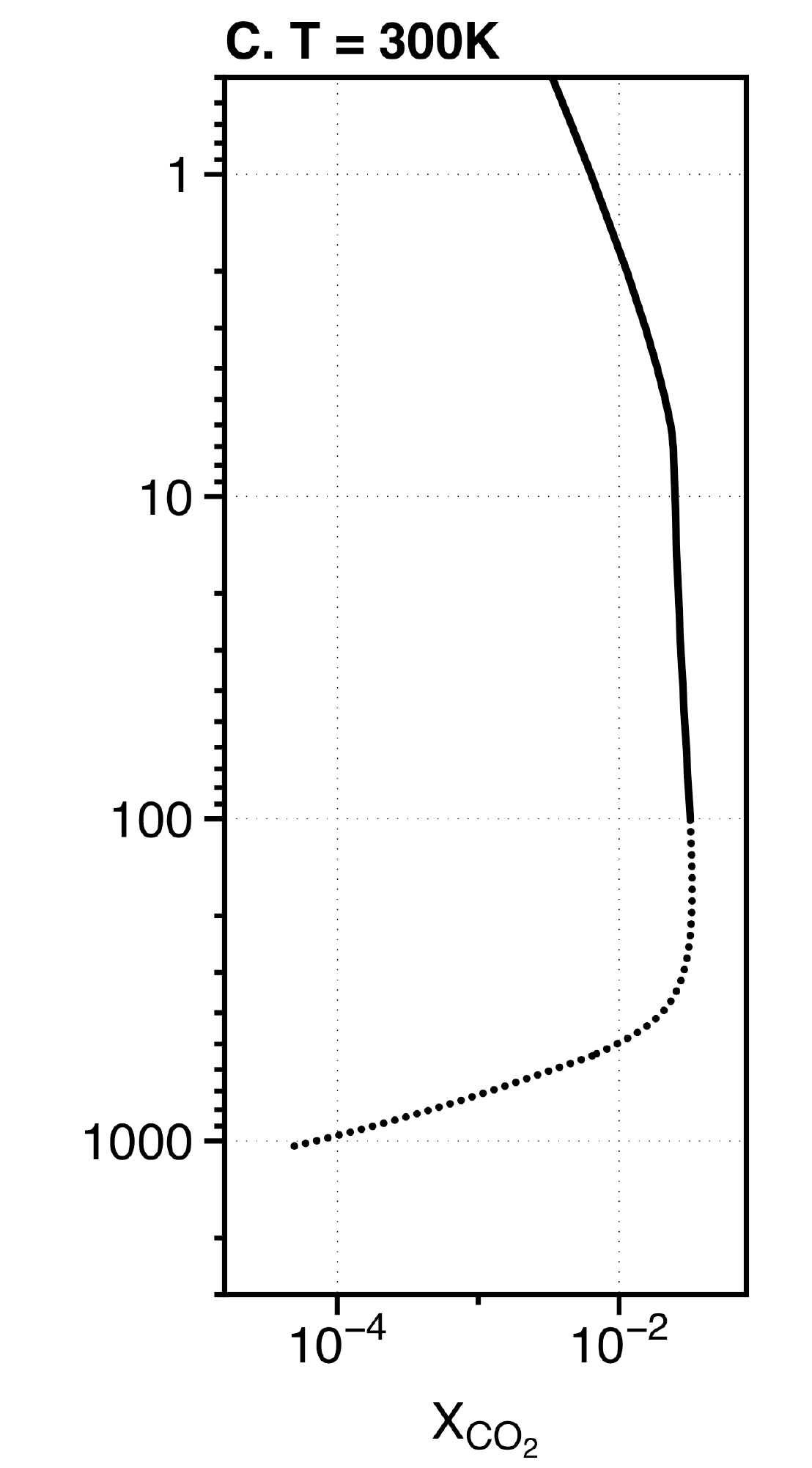}
    \includegraphics[width=0.24\linewidth]{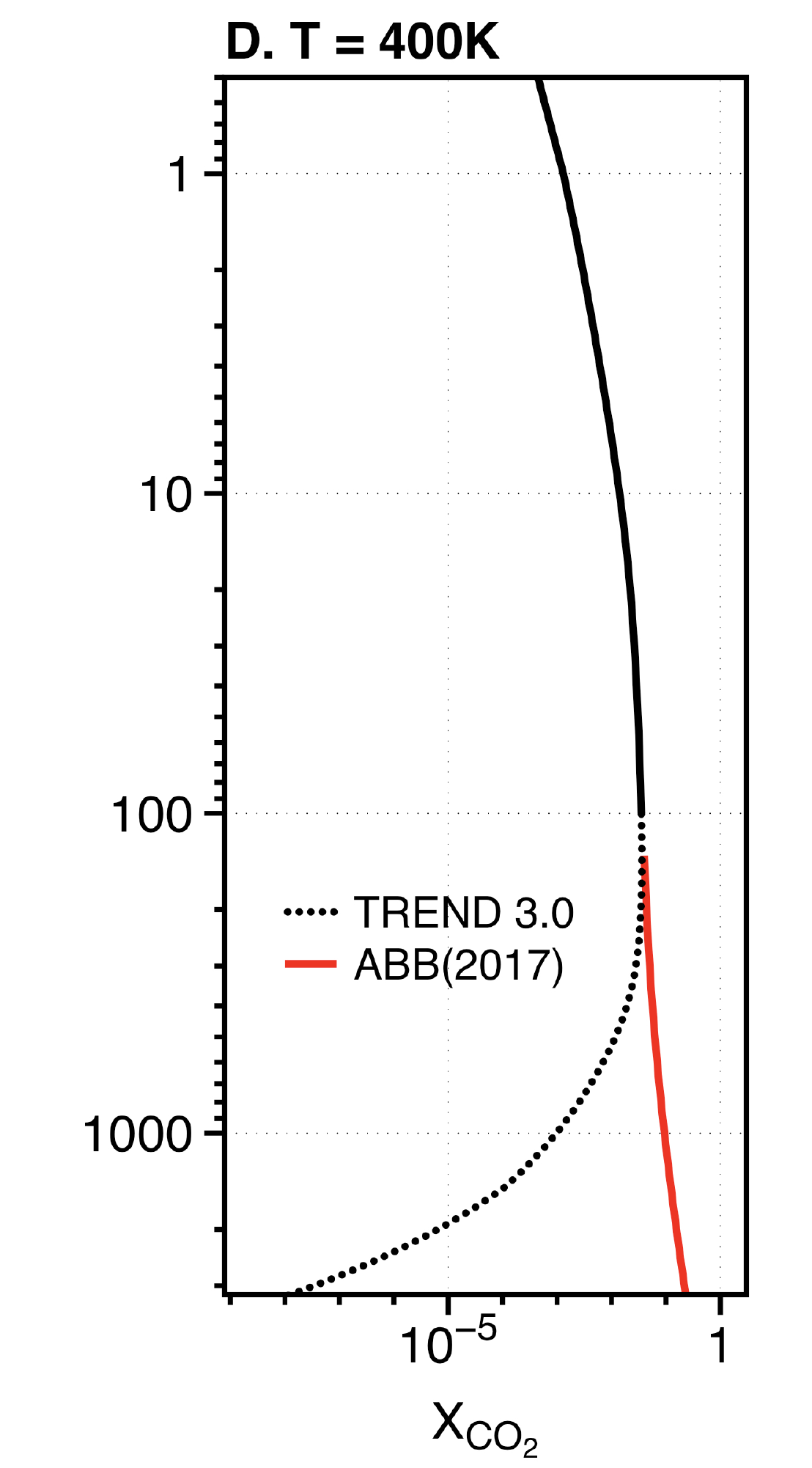}\\
    \includegraphics[width=0.24\linewidth]{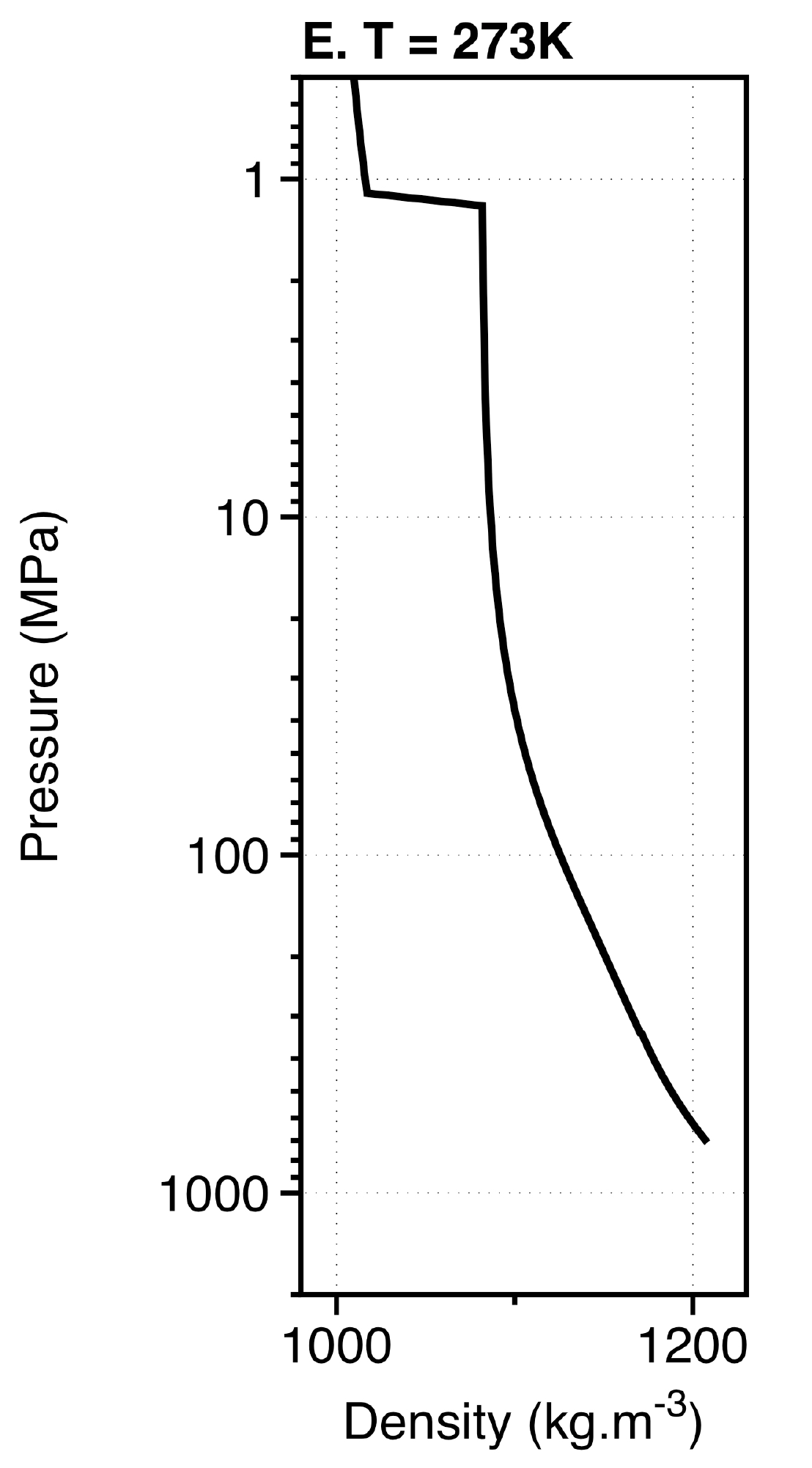}
    \includegraphics[width=0.24\linewidth]{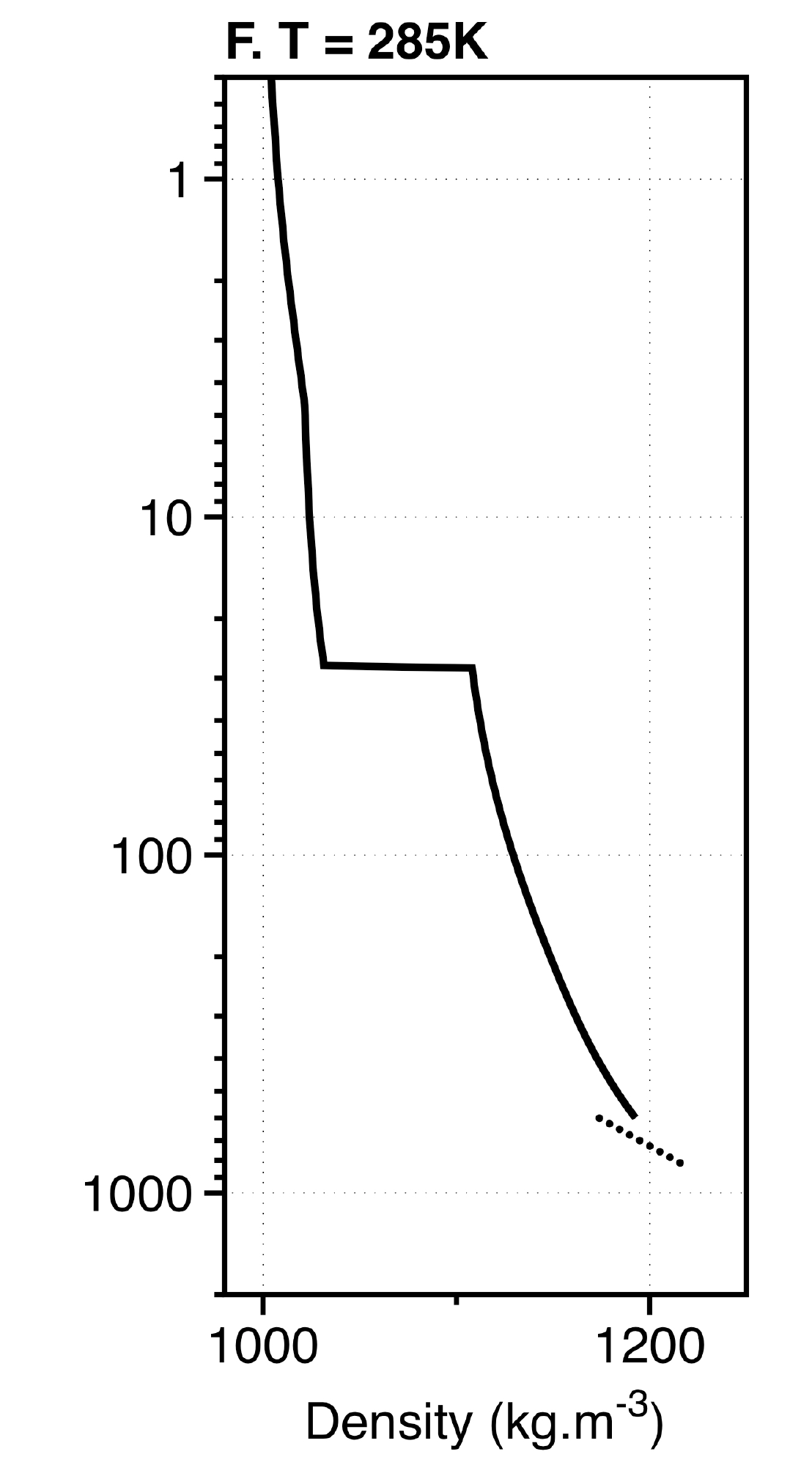}
    \includegraphics[width=0.24\linewidth]{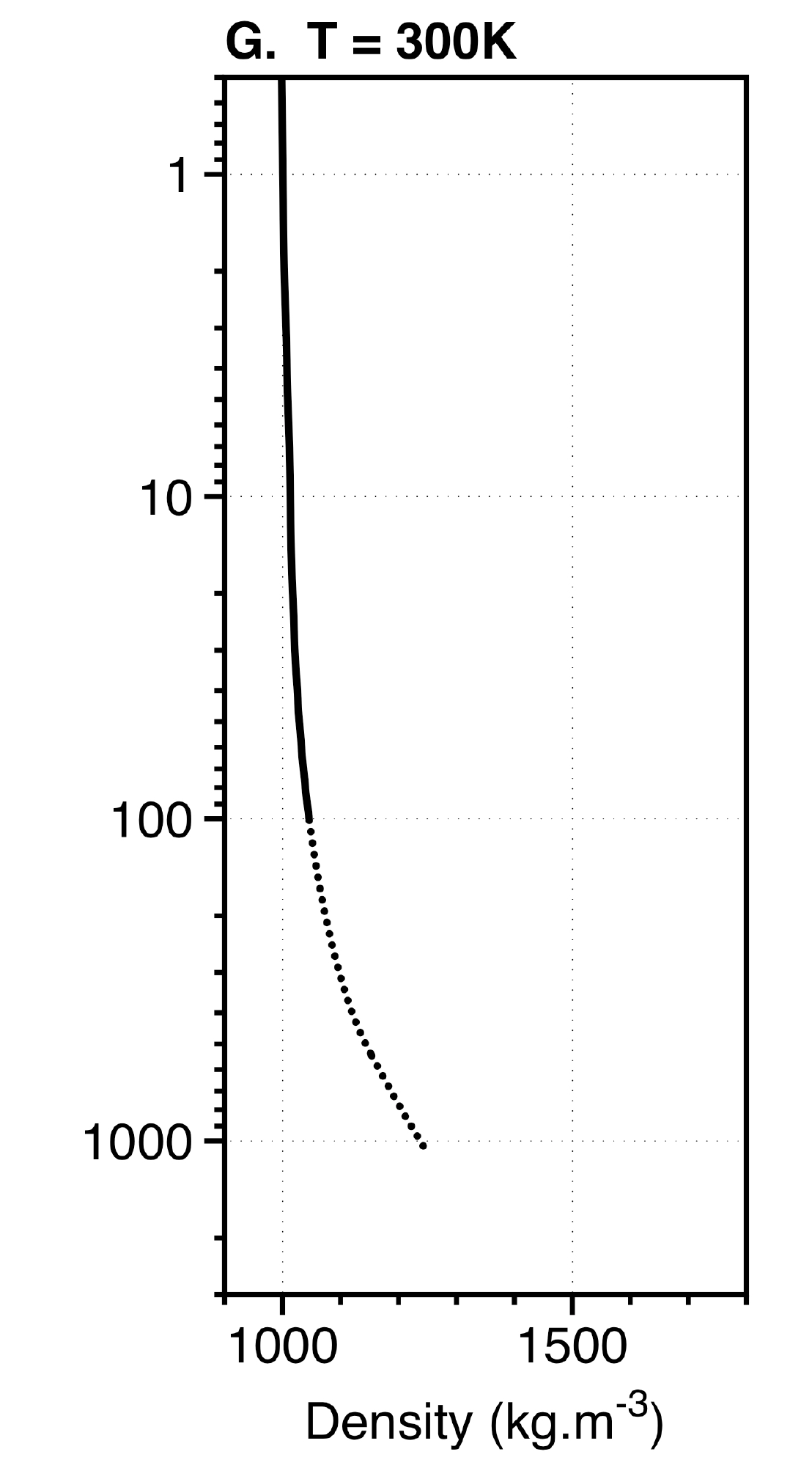}
    \includegraphics[width=0.24\linewidth]{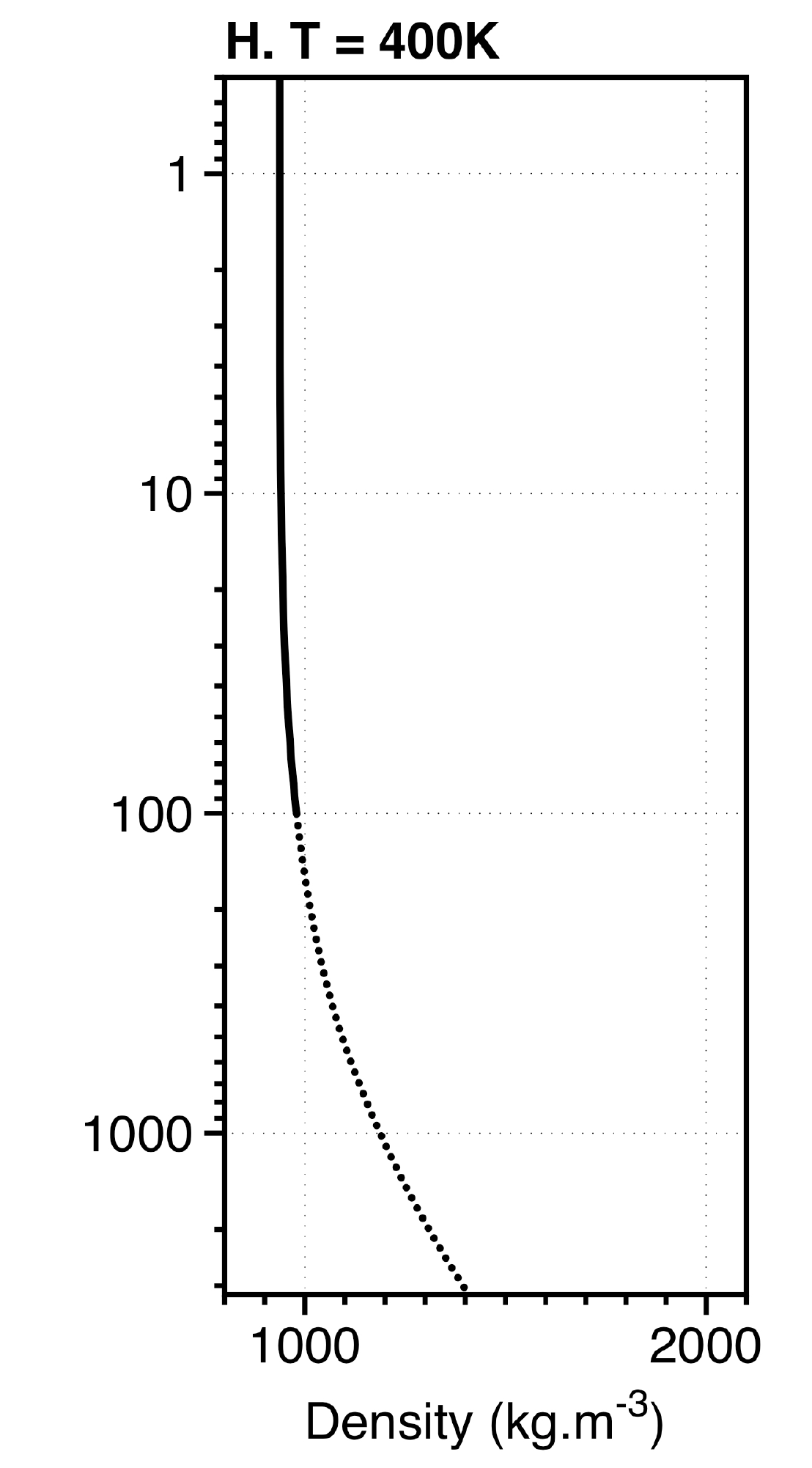}
    \caption{Molar fraction of CO$_2$ (A to D) and density (E to H) profiles of CO$_2$-saturated water-worlds for isothermal ocean profiles at T~=~273~K, 285~K, 300~K, and 400~K. Dotted lines denote pressures outside of the TREND 3.0 range of validity. The red line in panel D shows compositions interpolated from the experimental data of \citet{Abramson:2017ie} for T~=~373~K. Abrupt changes in composition and density denote phase transitions (see details in the text). Each profile is plotted until it reaches the pressure of formation of high-pressure water ice. Profiles inside the high-pressure ice phases are not shown.}

    \label{fig:profil}
\end{figure*}

The hydrospheres of water-worlds present a variety of structures, depending on temperature. Examples of these possible ocean structures are plotted in Fig. \ref{fig:profil}. In this figure, all of the profiles are plotted for an atmospheric pressure of 5 bar. We show in section \ref{sec:atm} that the amount of CO$_2$ in the atmosphere does not influence the results displayed in Fig. \ref{fig:profil} or Fig. \ref{fig:bilan}.

For temperatures between 273~K and and 279~K, CO$_2$ solubility increases with pressure in the liquid ocean until 1.2~MPa, where a thick CO$_2$ clathrate hydrate layer forms. Fig. \ref{fig:profil} A and E show an example of this profile for a T~=~273~K ocean isotherm. The phase transition from liquid water to clathrate is marked by a sudden jump in the CO$_2$ molar fraction and density. This clathrate layer is immediately in contact with the high-pressure water ice at P~=~710~MPa.

For ocean isotherms between 279~K and 294~K, water-world hydrosphere structures display two oceans: one on top of the clathrate layer and another one under the clathrate layer, immediately in contact with the high-pressure ice. Fig.~\ref{fig:profil} displays an example of this second type of hydrosphere structure for a 285~K ocean isotherm (panels B and F). Both density and composition profiles exhibit two phase transitions, one at 28~MPa and another at $\sim$600~MPa. The second phase transition occurs when the clathrate layer becomes unstable at higher pressures. This is due to the inversion of slope of L$_w$L$_c$H triple line (liquid water/liquid CO$_2$/clathrate hydrates) for pressures higher than 350~MPa (see Fig. \ref{fig:phasediagram}).

Clathrate hydrates contain up to 15~mol\% of CO$_2$, while the solubility of CO$_2$ in the global liquid ocean does not exceed $\sim$4~mol\%, making clathrate hydrates the main CO$_2$ reservoir for ocean isotherms T~$<$~294~K. The temperature of 294~K marks the stability limit of the clathrate phase: above it, clathrates are not stable at any pressure in the ocean. For T~=~300~K (Fig. \ref{fig:profil} C and G), the ocean is limited by the formation of high-pressure water ice. For the highest ocean temperature explored here, T~=~400~K (Fig. \ref{fig:profil} D and H), the liquid water region extends up to 3.2~GPa.

\subsection{Mass budgets of saturated reservoirs}

\begin{figure}
    \centering
      \includegraphics[width=1\linewidth]{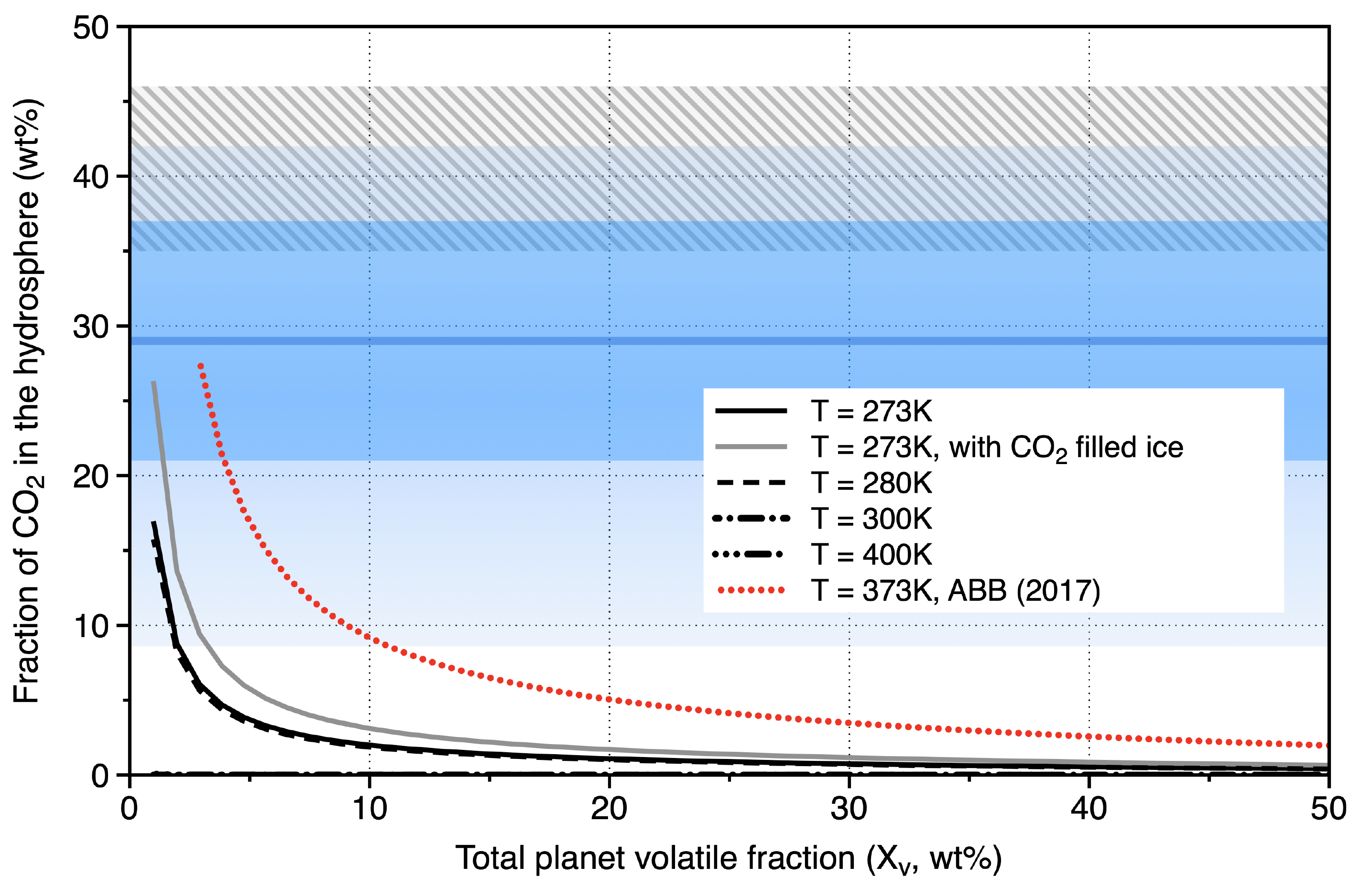}
  
    \caption{Hydrosphere saturated reservoir CO$_2$ storage capacity. The total mass fraction of CO$_2$ in the hydrosphere is plotted as a function of the bulk mass fraction of volatiles of the planet, for several oceanic isotherms. The low $X_v$ extreme of each curve is defined by the minimum volatile mass fraction at which the water-worlds develop a high-pressure water ice mantle. The red line adopts CO$_2$ solubility interpolated from the experimental data of \citet{Abramson:2017ie}. The grey line includes, in addition to the CO$_2$ dissolved in the ocean and trapped in clathrates, the possible contribution of the CO$_2$-filled ice (see \ref{sec_CO2filled}). For temperatures of 300~K and 400~K the CO$_2$ fractions are low ($<$0.01\%, see text) and the lines are nearly coincident with the horizontal axis of the plot. The range of CO$_2$ fractions shaded in blue shows the range of mass fraction detected in comets \citep{Mumma:2011jn, Ootsubo:2012cg}. The darker blue region encompasses the upper and lower quartile of the 17 comets sample, analyzed in \citet{Ootsubo:2012cg}, while the blue line shows the median value of this sample. The striped region shows the range of CO$_2$ fractions in low-mass protostellar envelopes \citep{Oberg:2011ev}.
    }
    \label{fig:bilan}
\end{figure}

We now evaluate the total CO$_2$ mass contained in the saturated reservoirs at a given temperature, using the compositional profiles from \S~\ref{sec:profil}. This mass is then divided by the total mass of the hydrosphere ($M_v$, including both CO$_2$ and water), to obtain the CO$_2$ fraction in the volatiles in the hydrosphere ($X_{CO_2}$, the vertical axis of Fig. \ref{fig:bilan}). We vary the total amount of volatiles accreted by our model planets from $X_v=$~1~wt\% to $X_v=$~50~wt\%. The main difference here between a planet that accreted 1~wt\% volatiles and another that accreted 50~wt\% is that the latter has a thicker high-pressure water ice mantle, considered here to be pure water ice. Consequently, the calculated CO$_2$ fraction in the volatiles of the planet's saturated reservoirs decreases as $X_v$ increases. The cases that do not allow the formation of the high-pressure water ice at the oceanic floor ($X_v\lesssim$~3~wt\% at 373~K and 400~K) are not displayed in the Fig. \ref{fig:bilan}.

Fig. \ref{fig:bilan} shows that, for cold oceans (i.e., sufficiently cold to form clathrates $<$~294~K), only planets that accreted low (X$_v$~$<$~2~wt\%) bulk fraction of volatiles could store cometary abundances of CO$_2$ in their saturated reservoirs. In this temperature regime, decreasing temperatures lead to the formation of thicker CO$_2$ clathrate layers (because clathrates of CO$_2$ are stable over a larger pressure range). Since clathrates can store more CO$_2$ per unit mass than the liquid ocean (see Fig. \ref{fig:profil}), the ability of the hydrosphere to store CO$_2$ in the saturated reservoirs increases as ocean temperature decreases. 
 
For T~$>$~294~K, the clathrate layer is not stable and CO$_2$ dissolved in liquid water becomes the dominant saturated reservoir. This leads to low CO$_2$ fractions in the hydrosphere, not exceeding 0.01~wt\% for our highest temperature isotherms (300~K and 400~K).

 \subsection{Contribution of CO$_2$-filled ice} 
 \label{sec_CO2filled}
 Between pressures of $\sim$~0.6~GPa and 1~GPa, the CO$_2$-H$_2$O system can form a phase called CO$_2$-filled ice \citep{Hirai:2010hpa, Bollengier:2013jl, Tulk:2014du, Massani:2017fr}. CO$_2$-filled ice corresponds to a compressed hydrate phase, where CO$_2$ molecules fill the water channels. Above 1~GPa, this phase dissociates to CO$_2$ ice and ice VI. The presence of CO$_2$-filled ice has been investigated between 80~K up to 277~K. \cite{Amos:2017ki} estimated the mass ratio between CO$_2$ and water molecules in this phase to be 41~wt\%. Using this result, we estimate the size of the potential CO$_2$ reservoir in CO$_2$-filled ice and its influence on the hydrosphere CO$_2$ fraction, plotted as a grey line in Fig. \ref{fig:bilan}. We find that CO$_2$-filled ice can make a modest contribution to the total CO$_2$ budget of the water-world. The presence of this phase adds between $\lesssim$~1~wt\% and $\sim$~10~wt\% to the total fraction of CO$_2$ in the hydrosphere, allowing habitable water-worlds with $X_v$ up to 3.5~wt\% to reach comet-like CO$_2$ abundances (compared to the limit of $X_v<$ 2~wt\% in the absence of CO$_2$ filled ice).

\subsection{Effect of CO$_2$ dissociation at high pressures and temperatures}

 As warned by \citet{Gernert:2016bc}, the computation of the solubility of CO$_2$ in water by TREND~3.0 becomes more and more uncertain above $\sim$100~MPa. Recent experiments of the CO$_2$-H$_2$O system of \citet{Wang:2016if}, \citet{Abramson:2017ie, Abramson:2018} and the \textit{ab initio} simulations of \citet{Pan:2016} show that CO$_2$ strongly dissociates and forms carbonic acid (H$_2$CO$_3$) at high pressures and temperatures (P~$\gtrsim$~1~GPa, T~$\gtrsim$~373~K). This leads to a much higher solubility of CO$_2$ at high pressures than  predicted by TREND~3.0 (or that could be predicted by any of the currently available equation of state of CO$_2$-H$_2$O system, see \citealt{Abramson:2017ie} for a detailed discussion).

In Fig. \ref{fig:profil} D we plot in red the solubility of CO$_2$ in the liquid water ocean for T~=~373~K provided by the interpolation of the experimental data of \citet{Abramson:2017ie}. The solubility of CO$_2$ at 400~K is expected to be very similar to that plotted for 373~K, because at pressures above 1~GPa the isotherms display very similar compositions \citep[see Fig. 7 and 8 in][]{Abramson:2017ie}. The solubility of CO$_2$  reaches up to $\sim$~20~mol\% at 3~GPa and 373~K (see the red line in Fig.~\ref{fig:profil} D). 

The high-pressure dissociation of CO$_2$ investigated by \citet{Abramson:2017ie} increases the CO$_2$ storage capacity of deep liquid oceans by more than 3 orders of magnitude in some cases (red line in Fig. \ref{fig:bilan}). For example, the CO$_2$ storage capacity goes from 0.01~wt\% CO$_2$ at 400~K without the high-pressure dissociation to more than 10~wt\% with. This highlights the effect  of uncertainties in the solubility of CO$_2$ at high pressures on the structure and maximal global saturated CO$_2$ reservoirs of water-worlds.
With the high-pressure dissociation of \citet{Abramson:2017ie}, the CO$_2$ storage capacity of the ocean at 373~K eclipses that of clathrates at temperatures below 294~K. We find that hot habitable water-worlds may store cometary abundances of CO$_2$ in their water-dominated layers only if they initially accreted less than 11~wt\% of volatiles by mass, when the high pressure dissociation of CO$_2$ is taken into account.

We expect that the impact of high-pressure dissociation on the CO$_2$ mass budgets calculated in Figure~\ref{fig:bilan} will diminish with decreasing ocean temperature. Oceans at lower temperatures are shallower, reaching the interface between the liquid water and high-pressure ice or clathrates at lower pressures. CO$_2$ dissociation decreases with decreasing pressures, leading to lower CO$_2$ solubilities for pressures $\lesssim$~1~GPa. Consequently, the red line in Fig. \ref{fig:bilan} is an upper limit on the contribution of saturated reservoirs to the total CO$_2$ budget of water-worlds.


\section{CO$_2$ content of excess reservoirs}
\label{sec:ResultsExcess}

For planets with high volatile mass fractions (more than 3.5~wt\% or 11~wt\%, depending on the temperature), the liquid water ocean, clathrates and CO$_2$ filled ice (if present) are insufficient to store comet-like amounts of CO$_2$. Here we propose possible {\it excess} reservoirs that CO$_2$ may form upon the saturation of the water-dominated phases: the atmosphere, liquid CO$_2$, CO$_2$ ice, and monohydrate of carbonic acid (H$_2$CO$_3$ $\cdot$ H$_2$O).

\subsection{Atmospheres of habitable water-worlds}
\label{sec:atm}

\begin{figure}[h!]
    \centering
    \includegraphics[width=1\linewidth]{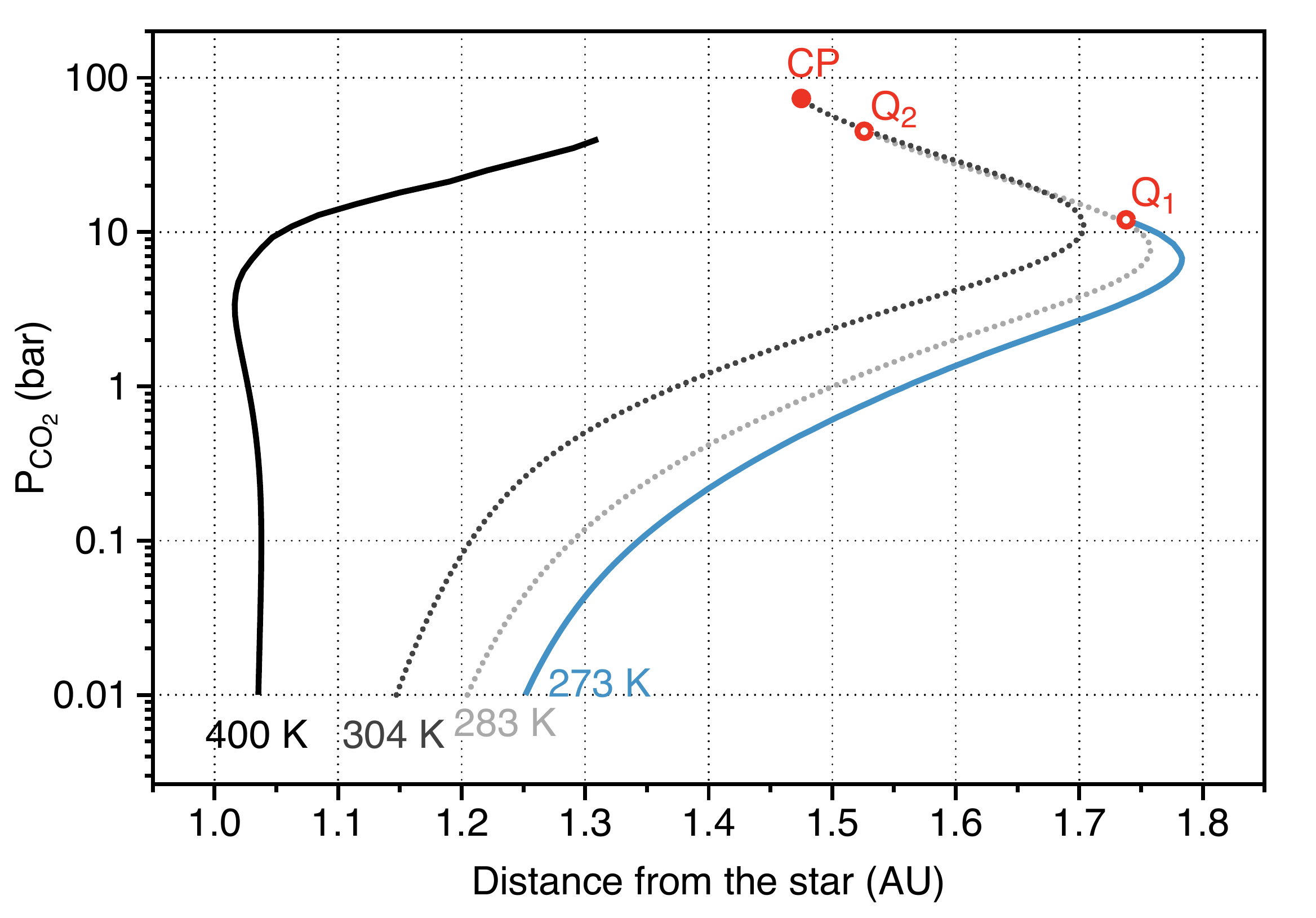}
    \caption{Boundaries of the habitable zone (HZ), as a function of the partial pressure of CO$_2$ in the atmosphere. For this figure the partial pressure of N$_2$ has been fixed to 1 bar and the surface albedo of the planet to 0.06, corresponding to the albedo of the Earth's ocean. Q$_1$ and Q$_2$ are quadruple points of CO$_2$ (same labels as in the Fig. \ref{fig:phasediagram}) and CP shows the critical point of CO$_2$.}
    \label{fig:HZ}
\end{figure}

Using the CLIMA 1D radiative-convective model we reproduce the boundaries of the water-world HZ, as a function of the partial pressure of CO$_2$ (first computed in \citet{Kitzmann:2015}, Fig. \ref{fig:HZ}). For visual reference, we have indicated locations of the critical point and the quadruple points Q$_1$ and Q$_2$ of the CO$_2$-H$_2$O phase diagram (Fig.~\ref{fig:phasediagram}). Limits of the HZ plotted here are uncertain due to the host star type, atmospheric 3D circulation, planet rotation rate, oceanic circulation, planet mass and possibly other factors \citep[e.g.]{Marshall:2007fj, Yang:2013gl, Kopparapu:2014, Kopparapu:2017, Turbet:2017cx, Ramirez:2018wn, Kite:2018}. We use these 1D simulations only to estimate the possible atmospheric masses of CO$_2$ of habitable water-worlds.

For a given distance from a Sun-like star, if the CO$_2$ content of the atmosphere exceeds the amount indicated by the black line, the surface temperature of this planet would exceed 400~K and the water-world would then be considered as uninhabitable. We stopped our simulations for the inner edge of the HZ at a partial pressure of CO$_2$ of 40~bar, because for higher pressures, the non-ideal behaviour of CO$_2$ must be taken into account and CLIMA treats CO$_2$ as an ideal gas at temperatures above 303~K.
For pure CO$_2$, the deviation from ideal pressure at 20 bar and $\sim$400~K is of the order of 5\%, while at 50~bar it reaches $\sim$20\% \citep[e.g.][]{Hu:2007ca, Duan:1992}. The addition of any other compounds (i.e. N$_2$ and H$_2$O) would only accentuate this error.

Fig. \ref{fig:HZ} indicates that above a partial pressure of $\sim$~10~bar, the distance from the star to the inner edge of the HZ sharply increases  \citep[see also][]{Kitzmann:2015}. An order-of-magnitude estimation --- $M_{CO_2}~=~P_{CO_2}4\pi R_p^2/g$, for Earth values of M$_p$, R$_p$ and g --- indicates that a CO$_2$ pressure of 100~bar represents a mass of the order of $5\times10^{20}$~kg, corresponding to $\sim$~0.01~wt\% of the total mass of the planet. Thus, CO$_2$ in the atmosphere would contribute at most $X_{CO_2}\lesssim$~1~wt\% to the total volatile budget of the water-world for the minimum $X_v\sim$1~wt\% that we consider, and even less for larger $X_v$.

We explored the influence of surface pressure on the CO$_2$ storage capacity of habitable water-worlds. We reproduced the results of Fig. \ref{fig:bilan} with an atmospheric pressure of  100~bar instead of 5~bar, finding no significant change ($\leq~10^{-3}$) in the final total CO$_2$ mass fractions of saturated reservoirs. 
Since we are considering a fully saturated case, the profiles of $X_{CO_2}$ with pressure throughout the ocean and clathrate layers (displayed in Fig.~\ref{fig:profil}) are unchanged; effectively, choosing a higher surface pressure of CO$_2$, only changes the low-pressure limit of the profiles. Since the atmosphere and ocean surface layer do not contribute significantly to the CO$_2$ storage capacity of habitable water-worlds, our results are insensitive to the choice of CO$_2$ atmospheric partial pressure.

\subsection{Condensation of liquid CO$_2$}

Liquid CO$_2$ is a potential excess reservoir for water-worlds near the outer edge of the HZ, with surface temperatures below the upper critical end point (UCEP) of the CO$_2$-H$_2$O system (304.5~K). At temperatures above the UCEP, liquid CO$_2$ will not condense at any pressure. Climate models of terrestrial planets have long recognized the possibility of condensation of CO$_2$ in Earth-like planet atmospheres and surfaces \citep{Kasting:1991, vonParis:2013hc}. These studies focused, however, on the limits imposed by the saturation vapor pressure on the amount of CO$_2$ in planetary atmospheres, and not on the liquid CO$_2$ condensate. The recent studies of \citet{Turbet:2017cx} and \citet{Levi:2017gv} investigated the possibility of condensation of liquid CO$_2$ and CO$_2$ clathrates at the poles of water-covered exoplanets.

For habitable water-worlds, condensation of liquid CO$_2$ at the surface is possible only if the partial pressure of CO$_2$ in the atmosphere is at least $P_{CO_2}>45$~bar and if the planet temperature lies between the second quadruple point (283~K) and UCEP (304.5~K) of the H$_2$O-CO$_2$ system. At temperatures below 283~K, CO$_2$ clathrates would form at the planet surface before liquid CO$_2$, assuming thermodynamic equilibrium. At temperatures above 304.5~K, CO$_2$ does not condense. Thus, surface liquid CO$_2$ oceans are stable in a rather narrow range of pressure and temperature conditions. 3D simulations \citep[e.g.,][]{Turbet:2017cx, Ramirez:2018wn} are thus needed to assess the long-term stability of liquid CO$_2$ as a CO$_2$ reservoir at the surface of water-worlds.

\begin{figure}[h!]
    \centering
    \includegraphics[width=1\linewidth]{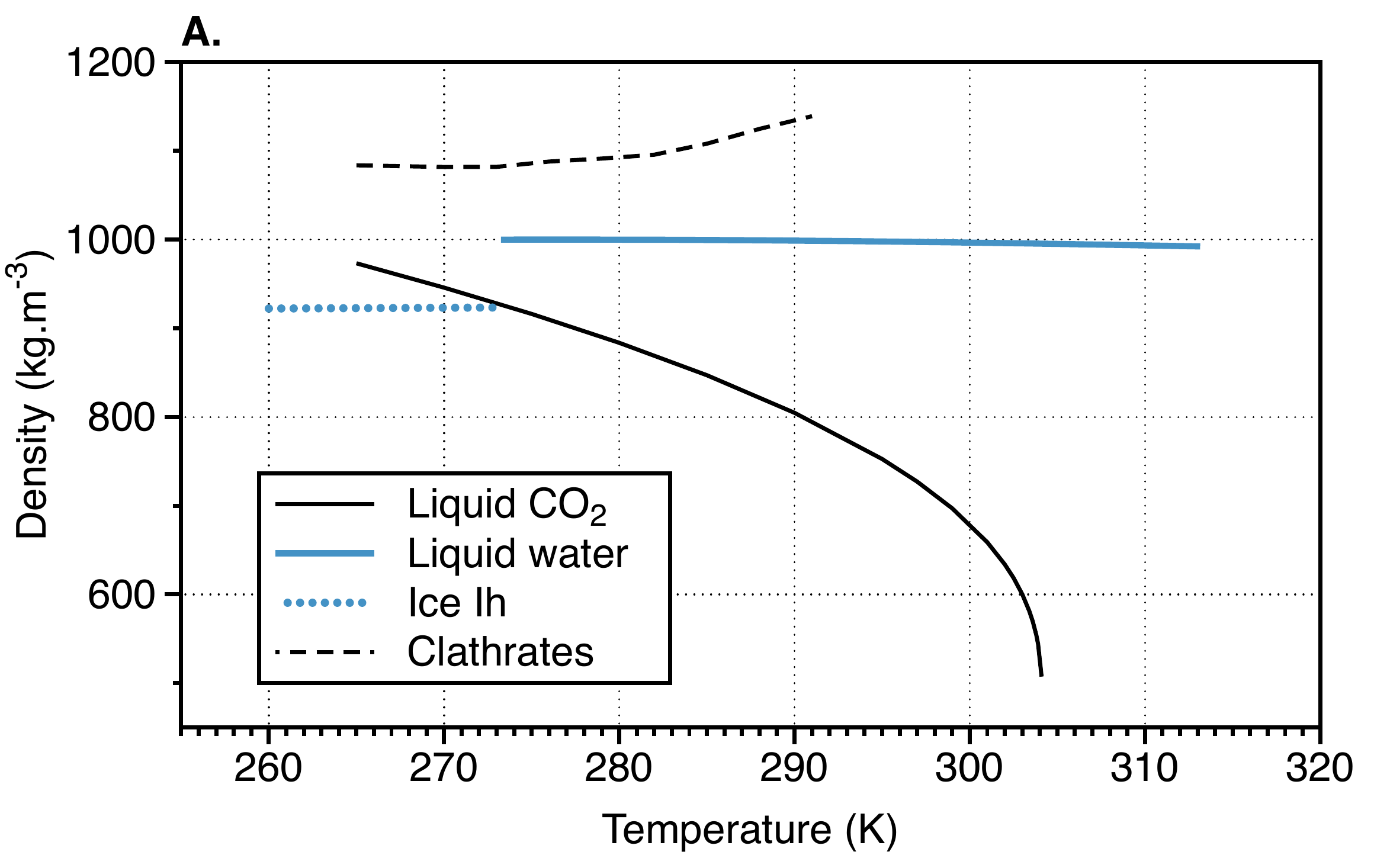}
    \includegraphics[width=1\linewidth]{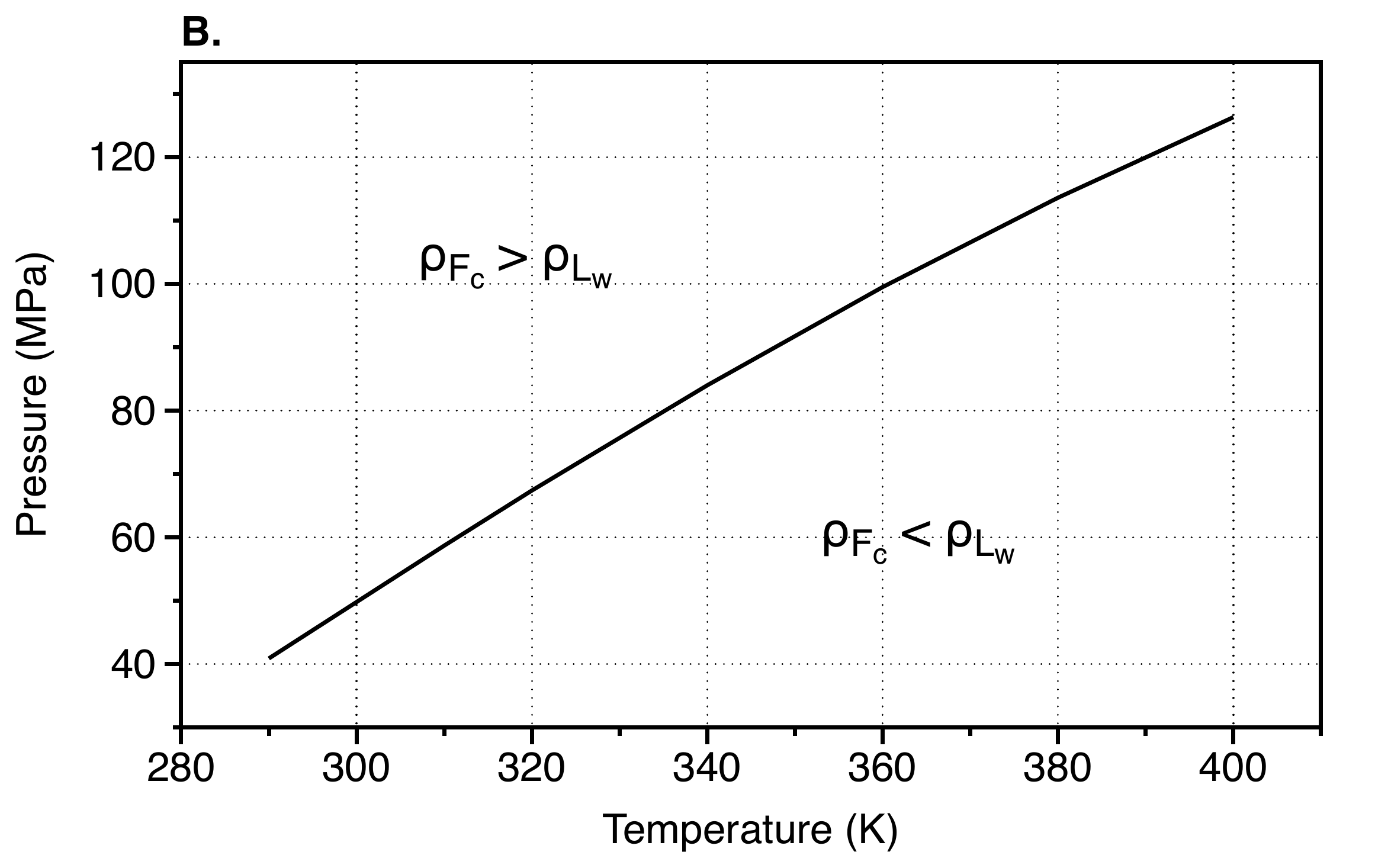}
    \caption{Comparison between densities of liquid water and liquid (T~$<$~304~K) or fluid CO$_2$ (T~$>$~304~K). Panel (A) compares the densities of liquid water, ice Ih, CO$_2$ clathrates and liquid CO$_2$ at low/surface  pressures (P~$<$~1~bar, at the saturation vapor pressures of each compound). The data plotted are from TREND 3.0 for liquid water, \citet{Choukroun:2007cx} for Ih water ice and \citet{Duschek:1990fc} for liquid CO$_2$. Panel (B) shows pressure as a function of temperature on the dividing line where the density of the water-rich liquid and the CO$_2$-rich fluid are equal (from TREND 3.0).}
    \label{fig:densityCO2water}
\end{figure}

 The density of liquid CO$_2$ varies considerably with temperature.  Fig.~\ref{fig:densityCO2water} shows that along its saturation line and for temperatures T~$>$~283~K, the density of liquid CO$_2$ will be lower than the density of water, allowing it to float on top of the liquid water ocean. For temperatures T~$<$~273~K, if CO$_2$ condenses before the formation of clathrates and in presence of surface water ice, then it would sink under the Ih ice layer, as described in \citet{Turbet:2017cx}. To determine the possible presence of large areas of liquid CO$_2$ on top the liquid water ocean, one would need to account for temperature variations across planet surface, heat redistribution and CO$_2$ transport in the ocean and the atmosphere. If a surface liquid CO$_2$ ocean is present, the atmospheric partial pressure of CO$_2$ would be at (or near) saturation.

At higher pressures (between 40 and 130~MPa, depending on temperature, Fig. \ref{fig:densityCO2water} (B)), the density of fluid CO$_2$ can be higher than the density of liquid water. If CO$_2$ liquid-liquid phase separation occurs deep in the water ocean, i.e. at pressures higher than the line displayed in Fig. \ref{fig:densityCO2water} (B), then the liquid CO$_2$ would be denser than the ambient water and would sink toward the oceanic floor. There, the liquid CO$_2$ would cross the stability domain of CO$_2$ ice (see Fig. \ref{fig:phasediagram}), freeze and sink in the high-pressure water ice mantle, as elaborated below (Section \ref{sec:CO2ice}).

\subsection{CO$_2$ ice as a main reservoir of CO$_2$ for habitable water-worlds}
\label{sec:CO2ice}

\begin{figure}
\centering
\includegraphics[width=1\linewidth]{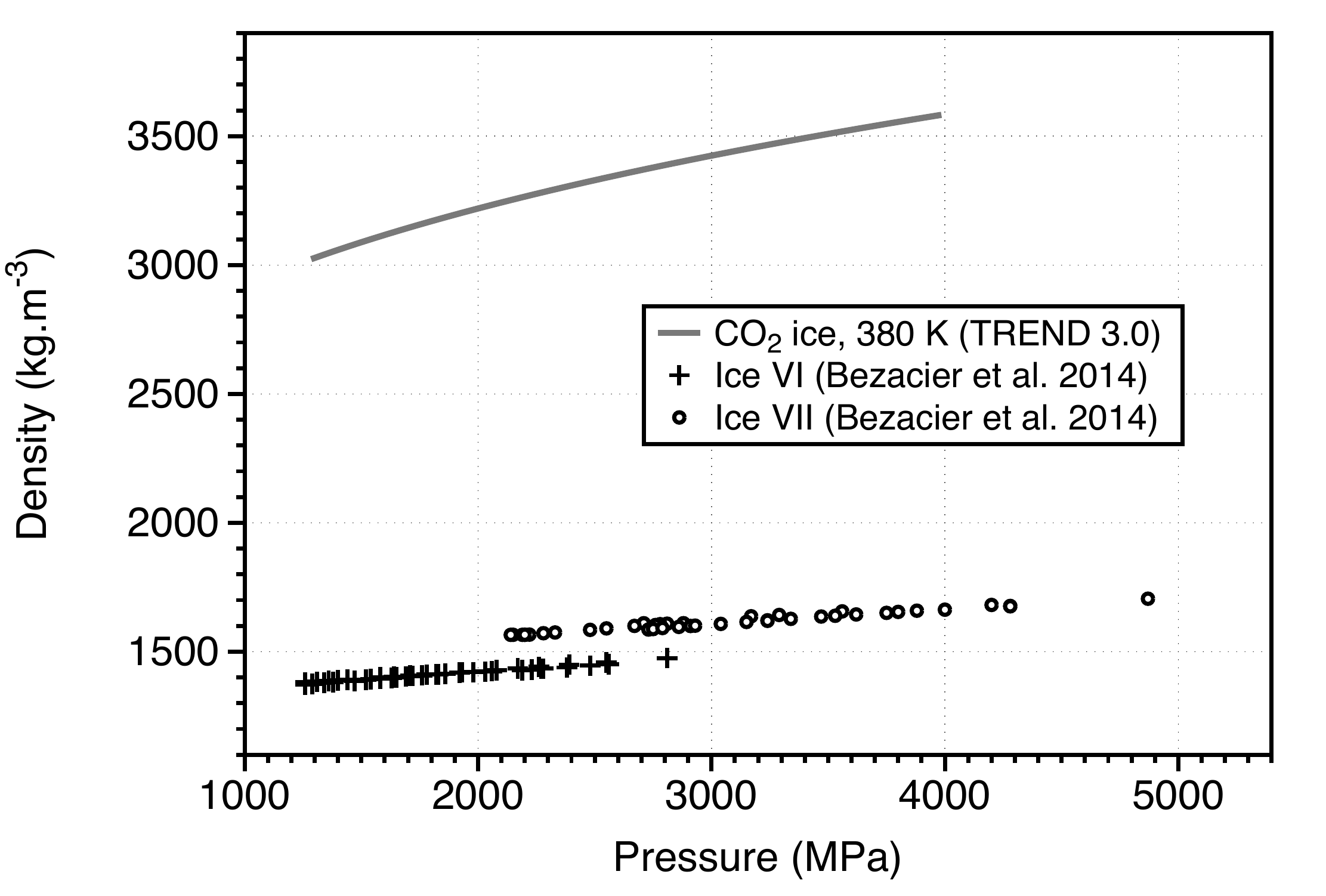}
\caption{Comparison between the density of CO$_2$ ice and high-pressure water ices. We use TREND 3.0 to obtain the density of CO$_2$ ice as a function of pressure for a 380~K isotherm. Points are experimental data of \citet{Bezacier:2014ii} for ice VI and ice VII, for temperatures ranging from 300~K to 340~K, and 300~K to 380~K, respectively.}
\label{fig:density}
\end{figure}

Water-worlds may encounter the conditions for CO$_2$ ice formation within their hydrospheres, as the planets evolve and cool. Whether the CO$_2$ ice forms before, during, or after the water-world's high-pressure water ice mantle freezes determines the initial formation location of the CO$_2$ ice layer (be it under, within or on top of the high pressure water ice mantle, see also Section \ref{sec:Discussion}). Because the density of CO$_2$ ice is always higher than the density of high-pressure water ices VI and VII (Fig \ref{fig:density}), CO$_2$ ice is stably stratified when buried under those layers. If the CO$_2$ ice is instead deposited on top of the high-pressure water ice later in the evolution of the planet, an unstable density stratification results, which could lead to gravitational Rayleigh-Taylor instabilities (and the eventual burial of CO$_2$ ice in the high-pressure water ice mantle). 

We estimate the timescales for the development of a Rayleigh-Taylor instability in CO$_2$ ice deposited at the surface of the high-pressure ice mantle. The timescale for the development of a Rayleigh-Taylor instability ($\tau_{RT}$) is the nominal time required to produce unit strain under a deviatoric stress of magnitude $g \times \Delta \rho \times  H $ \citep[e.g.][]{Turcotte:2014}:

\begin{equation}
    \tau_{RT}=\frac{13.04 \eta}{g \Delta \rho H },
    \label{eq:RT}
\end{equation}

\noindent where $\eta$ is the viscosity of the most viscous layer, $g$ is the local gravitational acceleration, $\Delta \rho$ is the difference in density between the two layers and $H$ is the length scale, set here to the thickness of the CO$_2$ layer.

The most viscous layer controls the development of a Rayleigh-Taylor instability. For temperatures below 343~K, CO$_2$ ice is in contact with ice VI. Laboratory measurements of the viscosity of ice VI at high differential stresses ($\gtrsim10^6$~Pa) give viscosities of the order of 10$^{13}$-10$^{14}$~Pa\,s \citep{Poirier:1981db, Sotin:1985, Durham:1996kq}. The viscosity of CO$_2$ ice (I) for similar differential stresses is of the order of 10$^{12}$- 10$^{13}$~Pa\,s. \citep{Durham:1999}. 

The extrapolation of the behavior of each material to planetary shear stresses must be handled with care. The creep behaviour at the low shear stresses ($\lesssim10^5$~Pa) relevant to planet interiors might be controlled by different mechanisms than those probed in experimental studies \citep{Durham:2001}. However, no experimental work to date has confirmed a change in creep behaviour for ice VI and CO$_2$ ice at decreasing shear stresses. Therefore, to estimate the viscosity of ice VI and CO$_2$ in planetary conditions, we adopt the shear stress dependence provided in \citet{Poirier:1981db} and \citet{Durham:1999}, respectively. We obtain upper estimates of 10$^{18}$~Pa\,s for ice VI and 10$^{15}$~Pa\,s for CO$_2$ ice \citep{Durham:2010bi}. Consequently, ice VI would control the formation of the Rayleigh-Taylor instability.

For temperatures above 343~K, CO$_2$ ice would be in contact with ice VII (Fig. \ref{fig:phasediagram}). To our knowledge, no experimental measurements of the viscosity of ice VII currently exist. The only theoretical study of the ice VII viscosity is found in \citet{Poirier:1982}, where the author examined the crystalline lattice of ice VII and concluded that ice VII should have a high viscosity because its crystalline structure does not favor the propagation of dislocations. Thus, ice VII is likely more viscous than the CO$_2$ ice and would control the formation of Rayleigh-Taylor instability at T~$>$~343~K.

\begin{figure}
    \centering
    \includegraphics[width=1\linewidth]{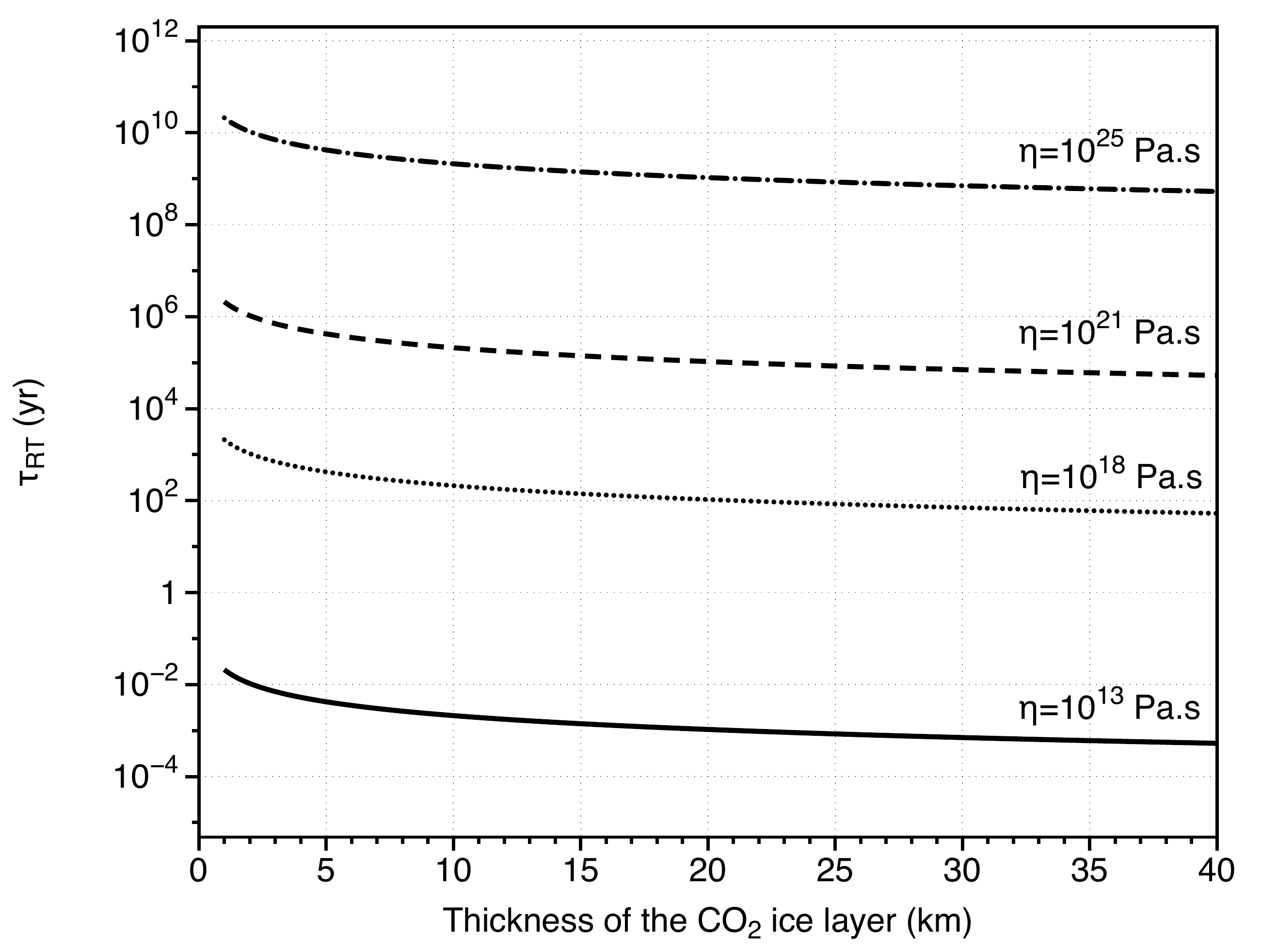}
    \caption{Timescales for the development of a Rayleigh-Taylor instability for a layer of CO$_2$ ice on top of the pure ice VI or VII. Due to the uncertainty of the viscosity of ice VI and VII for the planetary conditions, this figure displays a wide range of viscosities: $\eta=10^{13}$~Pa.s is the highest order of magnitude viscosity of ice VI experimentally measured by \citep{Poirier:1981db} and \citep{Durham:1996kq}; $\eta=10^{18}$~Pa.s is the extrapolation of these experimental measurements up to the planetary scales \citet{Durham:2010bi}. For a visual reference, we also show the same timescales for $\eta=10^{21}$~Pa.s, the viscosity of the Earth's mantle and for $\eta=10^{25}$~Pa.s, for which timescales of the development of Rayleigh-Taylor instability starts to be comparable to planetary ages.}
    \label{fig:RTinstability}
\end{figure}

Figure \ref{fig:RTinstability} shows timescales of the development of the Rayleigh-Taylor instability for several choices of $\eta$. The viscosity values span from the lowest viscosities experimentally measured for ice VI, to $\eta=10^{25}$~Pa \,s, where the timescales to form Rayleigh-Taylor instabilities start to be comparable to the planetary ages.

For ice VI, the timescale of the development of Rayleigh-Taylor instability is always short, less than 10,000~years. Unless ice VII viscosity is greater than $\eta=10^{25}$~Pa.s, CO$_2$ ice would also sink in the mantle in short timescales for temperatures higher than 343~K. We thus conclude that the formation of a CO$_2$ ice layer at any time during the thermal evolution of a water-world would lead to its burial inside of high-pressure ice mantle.

If CO$_2$ is buried as CO$_2$ ice inside of the high-pressure water ice mantle, the atmospheric CO$_2$ content would be decoupled from the total CO$_2$ content of the water-world. Consequently, future observations of habitable water-worlds' atmospheres would not provide the total CO$_2$ content of the planet but the CO$_2$ content of the atmosphere only. With models like the one presented in this study, it would be possible to estimate the amount of CO$_2$ dissolved in the global water ocean and trapped in clathrates, for an assumed interior temperature profile. However, the masses of CO$_2$ ice that might be trapped in the high-pressure water mantle are independent of these measurements and would be more challenging to constrain with future observations.

\subsection{Monohydrate of Carbonic Acid}

The H$_2$O-CO$_2$ phase diagram is poorly characterized at pressures above $\sim$1~GPa. A recent discovery is the formation of a carbon-bearing solid at high temperatures and pressures: the monohydrate of carbonic acid (H$_2$CO$_3$~$\cdot$~H$_2$O) \citep{Abramson:2017ie, Abramson:2018}.

\citet{Abramson:2017ie} identified the presence of a stable solid phase in the H$_2$O-CO$_2$ system for $P>4.4$~GPa. \citet{Abramson:2018} then determined that this phase consists of a monohydrate of carbonic acid, measured its crystalline structure and derived its density at $P=6.5$~GPa and $T=413.15$~K.

The temperatures and pressures at which  the monohydrate of carbonic acid has been observed correspond to the conditions in the  high-pressure ice layers of habitable water-worlds. The density of this solid is higher than the density of ices VI and VII (2194~kg.m$^{-3}$, \citet{Abramson:2018}, compare to Fig. \ref{fig:density}). Consequently, monohydrate of carbonic acid will remain buried in the high-pressure ice layers and coexist with CO$_2$ ice. Both of these solids would constitute an excess reservoir of CO$_2$ that would sequester the carbon dioxide away from the atmosphere.

The stability field of monohydrate of carbonic acid has not yet been mapped. 
It is still unclear if the monohydrate of carbonic acid (or other solids) form at lower pressures and temperatures (P~$<$~4.4~GPa and T~$<$~438.15~K, see \citet{Saleh:2016ew}). Any future detection of solid compounds in the H$_2$O-CO$_2$ system at these conditions would impact the interior models of water-worlds and icy satellites.


\section{Discussion}
\label{sec:Discussion}

\subsection{Planet Mass Dependence}

\begin{figure}
    \centering
    \includegraphics[width=1\linewidth]{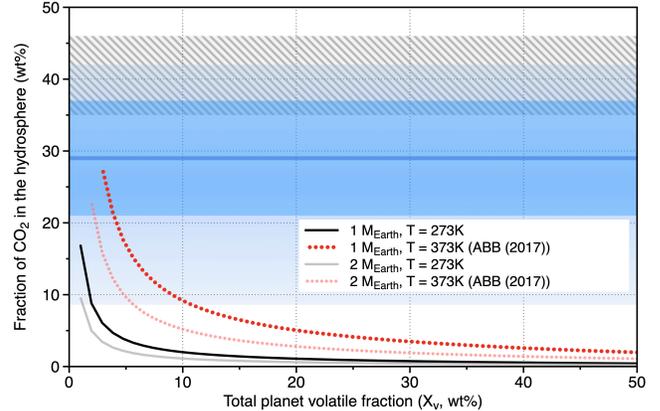}
    \caption{Dependence of hydrosphere saturated reservoir CO$_2$ storage capacity with $M_{rock}$. The total mass fraction of CO$_2$ in the hydrosphere is plotted as a function of the bulk mass fraction of volatiles of the planet, for oceanic isotherms of 273~K and 373~K. The $M_{rock}=1~M_{Earth}$ isotherms are the replication of the results displayed on the Fig.~\ref{fig:bilan}. ABB (2017) references the solubility obtained from the interpolation of the experimental data of \citet{Abramson:2017ie}. See the caption of Fig.~\ref{fig:bilan} for the description of shaded/stripped regions on the figure.}
    \label{fig:2Mearth}
\end{figure}

Our constraints on the CO$_2$ content of saturated reservoirs are more severe for planets with more massive rocky (iron and silicate) cores. For water-worlds with $M_{\rm rock}=2~M_{\rm Earth}$ (with Earth-like Fe/Si values), cometary compositions would be compatible with habitable surfaces if planets initially have less than 6~wt\% of volatiles by mass (see Fig. \ref{fig:2Mearth}), compared to the limit of 11~wt\% volatiles for planets with $M_{\rm rock}=1~M_{\rm Earth}$. More massive planets reach higher pressures in their hydrospheres due to their higher surface gravity. Consequently, they reach the stability field of high-pressure ice at lower depths. The ocean and the clathrate layers are thinner, and store less CO$_2$, when compared to lower mass planets with the same ocean temperature.

\subsection{Metal Core as another potential reservoir of CO$_2$}

In this work, we have focused on CO$_2$ reservoirs within the hydrospheres (water-rich outer layers) of water-worlds. Depending on the formation history of the planet, the rocky interiors of water-worlds offer another potential CO$_2$ reservoir.

Experimental studies show that CO$_2$, and more generally carbon, behaves as a siderophile element and is more likely to dissolve in the liquid iron than in silicate melts \citep[e.g.][]{Dasgupta:2013a}. Carbon solubility in iron-rich liquids increases with pressure and decreases with increasing temperature, extent of hydration and oxygen fugacity \citep{Dasgupta:2012, Dasgupta:2013a}. Previous studies have shown that Earth's core is constituted from 5~\% to 10~\% of light elements \citep{Birch:1964}, with C as a plausible candidate along with H, S, O and Si \citep{Poirier:1994}. Indeed, the iron core could be Earth's dominant carbon reservoir \citep{Bergin:2015in, Hirschmann:2016jn}. The exact composition in light elements of Earth's core is currently unknown, and the conditions for the C supply and the Earth's core formation continue to be subject to debate \citep[see the summary in][]{Dasgupta:2013_chap}. Likewise, the efficiency of the CO$_2$ entrapment in the iron-rich liquid, in the context of the formation of differentiation of water-worlds, is currently unexplored.

\subsection{  Entrapment of CO$_2$ as CaCO$_3$ }
\label{sec:CO2vsRocks}

On the Earth, carbonates -- (Ca, Mg, Fe)CO$_3$ -- are an important reservoir of CO$_2$. On Earth, carbonate minerals form when dissolved cations (such as Ca, Mg, Fe) produced by silicate weathering react with carbonate ions, CO$_3^{2-}$ in the ocean. In this work, since we focus on planets with sufficient water to form high pressure ice mantles, we have so far neglected reactions between the silicate rocks and liquid water \citep[following e.g.][]{Leger:2004fh, Selsis:2007jx, Fu:2010ch, Kitzmann:2015, Levi:2017gv} and thus have neglected carbonate minerals as a CO$_2$ reservoir on water-worlds. The formation of carbonates is impeded on a water-world because 1) the high pressure ice mantle creates a physical barrier between the ocean and silicates, 2) high pressures ($\gtrsim$~0.6~GPa) at the silicate surface leads to a stagnant lid tectonic regime with little renewal of fresh silicates for weathering \citep{Kite:2009kv} and 3) incorporation of salts into high-pressure water ice can efficiently remove salts from the liquid ocean \citep{Levi:2018hp}. In this section we elaborate further on the justification for this baseline assumption, and set an upper limit on the capacity of carbonates as a CO$_2$ storage reservoir on water-worlds. 

The thick high-pressure ice mantles of water-worlds may impede chemical exchanges between silicates and the liquid ocean on water-world planets. On water-worlds, liquid water oceans may be separated from silicate rock by high pressure ice mantles that are up to $\sim$~4000~km thick. In the case of Ganymede, \citet{Kalousova:2018} showed that the melting of water at the base of the high-pressure water ice mantle and the transport of this water by convection to the upper liquid water ocean is favored only for thin ice mantles ($<$~200~km) and high viscosities in the high-pressure ices. \citet{Kalousova:2018} found that the transport of molten water from the base of the high-pressure water mantle to the liquid water ocean shut down for a mantle thickness greater than 400~km. While such detailed studies have not yet been accomplished for water-worlds, this indicates a trend of weaker convection and therefore less chemical exchanges for thicker high-pressure ice mantles. 

On Earth, plate tectonics and volcanism continuously provide fresh silicates at the Earth's surface, which replenish the source of cations for the carbonate formation.
\citet{Kite:2009kv} show that, on water-worlds, once the post-accretional magma ocean has solidified, high pressures ($\gtrsim$~0.6GPa) at the water-silicate interface curtail the volcanism. The resulting tectonic regime for the planet is a stagnant lid.  Water-worlds thus have a limited reservoir of silicates that would be available to supply cations to the ocean. 

If salts are initially present in the water-world ocean, they will be removed on timescales of tens of millions to hundreds of millions of years. \citep{Levi:2018hp} described a mechanism that can pump salts out of water-world oceans, sequestering them inside of the high-pressure ice.

Despite the arguments above, it is still to be determined if the presence of a thick high-pressure water ice layer would \emph{fully} impede chemical exchanges between the liquid water ocean and silicates. 
Water-rock interactions could occur during the early stages of planet formation (before the high-pressure ice mantle forms). They may also potentially occur during the later stages of the planet's evolution, if the heat flux at the top of the silicate layer allows for the melting of high-pressure ices \citep{Noack:2016bh, Kalousova:2018}. 

We set an upper limit on the CO$_2$ storage capacity of carbonates in ocean-bearing water-worlds, following a similar approach to \citet{Kite:2018}. \citet{Kite:2018} estimate that in the most optimistic case (assuming the liquid water reacts efficiently with the silicates) liquid water would react with a layer of silicates at most $\sim 50$~km thick. These reactions would primarily form CaCO$_3$, as Mg and Fe cations are more to likely form silicates (see \citet{Kite:2018} for a detailed discussion). 
For the purpose of this calculation, we assume that the silicates are basalts (which are ubiquitous in the Solar System). Basalts contain 11.39~wt\% of CaO, so if all of the calcium reacts with CO$_2$ to form CaCO$_3$, it would results in an entrapment of  $\sim7\times10^{21}\left(R_{rock}/R_{\oplus}\right)^2~\mathrm{kg}$ of CO$_2$. This mass of CO$_2$ stored in the carbonates, when scaled to the total volatile mass of the planet, $M_v$, corresponds to:

\begin{equation}
X_{\rm CO_2}=11~wt\% \left( \frac{R_{rock}}{R_{\oplus}}\right) ^2 
\left( \frac{X_v^{-1} -1}{0.01^{-1}-1}\right) 
\left(\frac{M_{\oplus}}{M_{rock}} \right).
\end{equation}
\noindent 
For a water-world with $M_{rock}=1~M_{\oplus}$ and $R_{rock}=1~R_{\oplus}$, carbonates can store an additional $ X_{CO_2}=11~wt\%$ when $X_v =1~wt\%$, and only $X_{CO_2}=2~wt\%$ when $X_v = 5~wt\%$. 
Thus, carbonates may extend the storage capacity of CO$_2$ for water-worlds with total volatile contents $<$5~wt\%, but represent a negligible CO$_2$ reservoir for planets with more massive volatile envelopes. Consequently, while carbonates could increase the CO$_2$ storage capacity in the hydrosphere at the low $X_v<$5~wt\% (left-most) edge of  Figs. \ref{fig:bilan} and \ref{fig:2Mearth}, the effect of carbonates is negligible over the rest of these figures. Our main conclusions still hold: water-worlds with hydrospheres accounting for more than 11~wt\% of the total planet mass require additional  CO$_2$ reservoirs (beyond the liquid ocean, clathrates if present, atmosphere, and carbonates) to both accommodate cometary abundances of CO$_2$ and host a surface liquid water ocean.

\subsection{Implications for Water-World Habitability}

Our models indicate that, in a majority of cases, water-worlds with comet-like compositions will be too CO$_2$ rich to host a liquid water ocean on their surfaces. Indeed, for water-worlds with more than 11~wt\% of volatiles, the saturated (water-dominated) layers of the hydrosphere cannot store comet-like amounts of CO$_2$. Moreover, the 11~wt\% limit is already generous. Our habitable water-world models almost never reach the median comet composition $X_{CO_2}~\sim29$~wt\%. Our initial assumptions of an isothermal oceanic temperature profile and full saturation of CO$_2$ maximize the amount of CO$_2$ stored in the water-dominated saturated reservoirs. If water-worlds are to be habitable, the excess CO$_2$ (i.e. the CO$_2$ that could not be incorporated in the saturated reservoirs) must be stored away from the ocean and the atmosphere.

Whether, how much, and where liquid and/or solid CO$_2$ layers form depend on the evolution and accretion history and the global CO$_2$ fraction of the planet. To determine whether the excess CO$_2$ is more likely to degas in the atmosphere, form a liquid layer on top of the water ocean, form CO$_2$ ice or monohydrate of carbonic acid, one would need to model the post-accretional cooling of a water-world's steam envelope from its initial hot state and determine at which pressures the CO$_2$-H$_2$O mixture saturates in CO$_2$. If this saturation is reached at low pressures (i.e. below the line of \ref{fig:densityCO2water} (B)), condensed CO$_2$ is likely to float on the surface of a water ocean and/or evaporate in the atmosphere. Consequently, a potential outcome for the evolution of water-worlds could be a hot hydrosphere consisting of an extended steam envelope that transitions from vapor to super-critical fluid, to plasma at greater and greater depths \citep[as has been considered by, e.g.,][]{Kuchner:2003, Rogers:2010fv, Nettelmann:2011dh, Lopez:2012hi}.
At high pressures, where CO$_2$ fluid is denser than water, the excess CO$_2$ would condense, sink and precipitate as ice or as monohydrate of carbonic acid. 
Our results therefore indicate that further modeling of water-world cooling and formation will be crucial for determining how much CO$_2$ each of these excess reservoirs could store and whether cometary composition water-worlds are likely to be habitable.

Previous studies \citep[e.g.][]{Kitzmann:2015, Levi:2017gv, Turbet:2017cx, Ramirez:2018wn} have implicitly assumed that water-worlds will form liquid water oceans if they're at an appropriate distance from the star and then derive an amenable CO$_2$ flux from the interior or CO$_2$ partial pressure in the atmosphere. Our work shows that the condensation of the high-pressure water ice layer as well as a liquid water ocean is not a given. Detailed 3D GCM modeling of water-worlds with liquid water oceans are inapplicable if water-worlds generically never manage to cool sufficiently to form liquid water oceans in the first place. Indeed, there is a tension between the likely comet-like compositions of volatiles accreted by water-worlds and the amount of CO$_2$ that can be accommodated in the water-world structures modeled in previous works. It is possible that the sequestration of solid CO$_2$ in the high-pressure ice mantle, as we have proposed in \S~\ref{sec:CO2ice}, could save the habitability of water-world exoplanets, but detailed models of the post-accretional cooling of water-worlds are needed to demonstrate this.

\subsection{Effect of equations of state uncertainties}

TREND~3.0 (and other cubic equations of state, when coupled to an appropriate clathrate model, i.e. \citealt{Gasem:2001, Soave:1972, Sloan:2007cl}) can accurately set an upper limit on the mass of CO$_2$ that could be stored in the water-dominated layers of ocean-bearing water-worlds at the outer edge of the HZ. For isotherms T~$<$~283~K, the depth of the ocean is limited by the formation of clathrates to isostatic pressures not exceeding 100~MPa. Experimental data are widely available for pressure-dependant CO$_2$ solubilities in these relatively shallow surface oceans. 

For ocean temperatures T~$>$~294~K, water-worlds have deep oceans and do not form clathrates. Current thermodynamic models \citep[including cubic equations of state, and TREND 3.0, for a review see][]{Abramson:2017ie}, strongly underestimate the amount of CO$_2$ dissolved in the deep liquid water ocean. This is likely due to the dissociation of CO$_2$ at high temperatures and pressures, that has been investigated only recently \citep[i.e.][]{Pan:2016, Abramson:2017ie} and is not included in the current equations of state for CO$_2$-H$_2$O system. Accounting for the dissociation of CO$_2$ at high pressure and high temperature greatly influences for the total CO$_2$ budget of saturated reservoirs. In Fig. \ref{fig:bilan}, accounting for this effect adds more than three order of magnitude in the total CO$_2$ storage capacity. Models that correctly reproduce the high solubility of CO$_2$ at high pressure and high temperature demonstrated by the recent experimental data are needed to assess the habitability of water-worlds.

Moreover, the phase boundaries of monohydrate of carbonic acid are currently poorly constrained \citep{Wang:2016if, Abramson:2018}. The experimental work of \citet{Wang:2016if} shows that this solid could form at high pressures ($>$~3.5~GPa) and high temperatures (1773~K), where neither high-pressures ice or CO$_2$ ice yet condense. Therefore, the accumulation of the monohydrate of carbonic acid could start at the early stages of water-world post-accretional evolution, removing CO$_2$ from a supercritical envelope. Detailed models of the cooling of water-worlds and the formation of such a reservoir are needed to understand if the removal of CO$_2$ by the precipitation of monohydrate of carbonic acid would be sufficient to lead to an efficient cooling of water-worlds and a subsequent condensation of liquid water and high-pressures phases of water ice. These models would necessitate  further constraints on the location of the monohydrate of carbonic acid phase limit in pressure-temperature-composition space.

We hope that this paper will motivate new experimental and computational studies to further explore the phase diagram of the CO$_2$-H$_2$O system, specifically the  dissociation of CO$_2$ in water at pressures above 1~GPa and the phase limit of the newly discovered solid monohydrate of carbonic acid.


\section{Conclusions}
\label{sec:Conclusion}
We model hydrosphere structures of water-worlds. We use the TREND 3.0,  a state-of-the-art equation of state widely used by the carbon capture and storage community, to determine the maximum amount of CO$_2$ dissolved in water and trapped in clathrates hydrates, as a function of temperature and pressure. We assume an isothermal profile in the liquid water and clathrates and an adiabatic profile in the high-pressure water ice mantle.

We determine that the atmosphere, ocean and clathrate layer can not be the main CO$_2$ reservoir on habitable (i.e., surface ocean-bearing) water-worlds that accreted more than 11~wt\% volatiles during their formation.
Water-worlds that accreted a smaller mass fraction of volatiles could potentially store comet-like amounts of CO$_2$ in their saturated reservoirs, depending on their temperature profile. Even then, in our models, the saturated hydrospheres of habitable water-worlds almost never reach the median comet composition of $X_{CO_2}~\sim0.29$~wt\%.

 If the excess CO$_2$ is not sequestered away from the atmosphere, habitable zone water-worlds may be unable to cool sufficiently from their post accretional hot state to condense liquid water oceans. The current paradigm of habitable-zone water-worlds as condensed, super-Ganymedes should be expanded. Depending on their post-accretional cooling history, we may be more likely to observe habitable zone water-worlds in hot uncondensed states with supercritical steam envelopes. 

We stress that extrapolations of current equations of state to high pressures and high temperatures (P~$>$~100~MPa, T~$>$~400~K) are unable to correctly predict the solubility of CO$_2$ in water. This is due to the dissociation of CO$_2$ at high pressures and high temperatures and is an ongoing research frontier in material science. Our work demonstrates that the dissociation of CO$_2$ has crucial implications for the habitability of water-worlds at the inner edge of the habitable zone.

Unless the C entrapment in the iron core was efficient during the accretion of water-worlds, the largest potential reservoir of CO$_2$ in the hydrospheres of habitable water-worlds is likely to be CO$_2$ ice and  monohydrate of carbonic acid, trapped in the high-pressure water ice mantle. Consequently, the atmospheric composition of an ocean-bearing water-world does not necessarily reflect the total mass of volatiles accreted during the formation of the planet, nor the relative proportions of CO$_2$ and H$_2$O in the hydrosphere. 

More detailed modeling of the  post-accretional cooling of water-worlds is needed to determine whether CO$_2$ ice burial could allow water-worlds to have liquid water oceans or whether the evolution of the planet would generically lead to too much atmospheric CO$_2$ for the planets to be habitable. 

\section*{Aknowledgements}
We would like to thank R. Span for providing us with the TREND 3.0 software, S. Domagal-Goldman R. Ramirez and R. Kopporapu for their help and for providing us with ATMOS, and E. Kite for insightful discussions about water-worlds.

\end{document}